\documentclass[10pt]{article} 
\usepackage{amssymb}
\usepackage{amsmath}
\usepackage{amsthm}
\usepackage{latexsym} 
\usepackage[dvips]{epsfig}
\usepackage{eufrak}
\usepackage{mathrsfs}

\theoremstyle{plain}
 
\newtheorem{lemma}{Lemma}
\newtheorem{theorem}{Theorem}
\newtheorem{assumption}{Assumption}
\newtheorem*{conjecture}{Conjecture}

\newtheorem*{definition}{Definition}

\def\O{\mathcal{O}}

\def\d{\mbox{d}}

\def\chiwt{\widetilde{\chi}}

\def\Dwt{\widetilde{D}}
\def\rwt{\widetilde{r}}

\font\SYM=msbm10 
\newcommand{\Real}{\mbox{\SYM R}}
\newcommand{\Complex}{\mbox{\SYM C}}

\newcommand{\Euclidean}{\mbox{\SYM E}}
\newcommand{\Integer}{\mbox{\SYM Z}}

\newcommand{\TT}[3]{T_{#1 \phantom{#2} #3}^{\phantom{#1} #2}}

\begin{document}


\title{\textbf{Time asymmetric spacetimes near null and spatial infinity. I. Expansions of developments of conformally flat data.} }

\author{Juan Antonio Valiente Kroon \thanks{E-mail address:
 {\tt jav@ap.univie.ac.at}} \\
 Institut f\"ur Theoretische Physik,\\ Universit\"at Wien,\\
Boltzmanngasse 5, A-1090 Wien,\\ Austria.}

\maketitle

\begin{abstract}
The conformal Einstein equations and the representation of spatial
infinity as a cylinder introduced by Friedrich are used to analyse the
behaviour of the gravitational field near null and spatial infinity
for the development of data which are asymptotically Euclidean,
conformally flat and time asymmetric. Our analysis allows for initial
data whose second fundamental form is more general than the one given
by the standard Bowen-York Ansatz. The Conformal Einstein equations
imply, upon evaluation on the cylinder at spatial infinity, a
hierarchy of transport equations which can be used to calculate
asymptotic expansions for the gravitational field in a recursive
way. It is found that the solutions to these transport equations
develop logarithmic divergences at certain critical sets where null
infinity meets spatial infinity. Associated to these, there is a
series of quantities expressible in terms of the initial data
(obstructions), which if zero, preclude the appearance of some of the
logarithmic divergences. The obstructions are, in general, time
asymmetric. That is, the obstructions at the intersection of future
null infinity with spatial infinity are in general different from
those obtained at the intersection of past null infinity with spatial
infinity. The latter allows for the possibility of having spacetimes
where future and past null infinity have different degrees of
smoothness. Finally, it is shown that if both sets of obstructions
vanish up to a certain order, then the initial data has to be
asymptotically Schwarzschildean to some degree.
\end{abstract}

\textbf{Pacs: 04.20.Ha, 04.20.Ex}

\section{Introduction}
The \emph{regular initial value problem near spatial infinity}
introduced by Friedrich in \cite{Fri98a} has proved an invaluable tool
for the understanding of the behaviour of the gravitational field in
the region of spacetime near spatial and null infinities. This
setting, relying entirely on the properties of conformal structure, is
such that the equations and data are regular with spatial and null
infinities having a finite representation with their structure and
location known \emph{a priori}. In order to keep the amount of
computations at bay, Friedrich's original analysis was restricted to
time symmetric initial data sets which admit a smooth conformal
compactification at infinity. Central to his discussion, laid a
representation of spatial infinity as a cylinder on which the
Conformal Einstein equations fully reduce to a set of transport
equations. The crucial feature of these equations is that through its
solutions it is possible to relate properties of the spacetime at null
infinity with properties of the initial data sets, effectively
allowing us to identify the parts of the initial data responsible for
the non-smoothness of null infinity. In particular, in \cite{Fri98a} a
regularity condition was formulated, that if satisfied, would preclude
the appearance of a certain type of logarithmic divergences in the
solutions of the transport equations at the sets where null infinity
intersects spatial infinity\footnote{This condition is expressed in
terms of the Cotton-York (Bach) tensor and its derivatives to all
orders. More precisely,
\[
\mathcal{S}(D_{i_1}\cdots D_{i_q}B_{jk})(i)=0 \mbox{ for } q=0,1,\ldots,
\]
where $D_i$ denotes the covariant derivative of the metric on the
initial hypersurface, $B_{jk}$ is its Cotton tensor, and $\mathcal{S}$
denotes the operation of taking the symmetric, tracefree part of a
tensor.}. Because of the underlying hyperbolic nature of the Conformal
Einstein equations, it is to be expected that the logarithmic
divergences would propagate along the generators of null infinity and
leave an imprint on various radiative properties of the spacetime, the
decay of the components of the Weyl tensor among them.

The solutions of the transport equations at spatial infinity can be
calculated by means of what is essentially a decomposition of the
unknowns in terms of spherical harmonics. The feasibility of this
approach was explored in \cite{FriKan00}. Leaving aside the
calculational complexities, this approach is completely algorithmic,
and hence, entirely amenable to an implementation in a computer
algebra system. In \cite{Val04a} this idea was put in practice for
conformally flat initial data sets which are conformally flat. This
family of initial data sets would satisfy trivially the regularity
condition found in \cite{Fri98a}. The calculations performed by means
of scripts written in the computer algebra system {\tt Maple V}
revealed that, generically, further logarithmic divergences would
arise. Associated with these divergences there is a hierarchy of
quantities written entirely in terms of the initial data
(\emph{obstructions}), that, if set to zero, would eliminate some of
the logarithmic divergences from the solutions of the transport
equations. Furthermore, an initial data set for which the obstructions
vanish up to a certain order, $p_*$, would be found to be asymptotic
Schwarzschildean to a certain order $s_*$ which would depend on $p_*$
---i.e. $s_*=s_*(p_*)$. This leads to conjecture that the only time
symmetric, conformally flat initial data sets yielding a development
admitting a conformal compactification are the Schwarzschildean ones.

Stationary (vacuum) spacetimes, and in particular the static ones, are
known to admit a smooth null infinity ---see
\cite{DamSch90}\footnote{Note, however, that the original proof had a
gap which was filled in \cite{Dai01b}.}. In order to be able to
consider static solutions one has to move away from the class of
conformally flat initial data sets as the Schwarzschild solution is
the only static solution having slices which are conformally flat near
infinity. In \cite{Val04c} the calculations carried out in \cite{Val04a} were
extended to the case of non-conformally flat spacetimes which
satisfy Friedrich's regularity condition. As expected, the
calculations rendered a generalised hierarchy of obstructions, which, if
satisfied, imply asymptotic stationarity to a certain order. Thus, it
seems that time symmetric initial data sets would render developments
admitting a null infinity if and only if they are asymptotic
static. Providing a proof of this conjecture would constitute a
remarkable feat.

Why so much interest in spacetimes admitting a smooth null infinity?
The driving force of the programme initiated in \cite{Fri98a} is to
verify the so-called \emph{Penrose proposal}. Loosely speaking,
Penrose's proposal states that the gravitational field of spacetimes
describing isolated systems ---e.g. two coalescing black holes--- should
satisfy in its asymptotic region the \emph{Peeling
behaviour}. That is, the components $\widetilde{\Psi}_n$ of the Weyl
tensor (where here we are using the standard Newman-Penrose notation)
should decay along the generators of future oriented light cones as
\begin{equation}
\widetilde{\Psi}_n=\O\left( \frac{1}{\overline{r}^{5-n}} \right), \quad n=0,\ldots 4,
\end{equation}
where $\overline{r}$ is an affine parameter along the generators of
the light cones. As it was already pointed out in \cite{Val04b},
spacetimes with obstructions like the ones found in
\cite{Val04a,Val04c} cannot be expected to peel. Thus, the question is
now how restrictive is Penrose's proposal when discussing systems of
physical interest? In any case, it should be noted that non-static
spacetimes satisfying the peeling behaviour do exist. These have been
constructed by Chru\'sciel \& Delay \cite{ChrDel02} from initial data
sets which are asymptotic Schwarzschildean outside a compact set. The
initial data sets have been obtained by using a refinement of the
deformation techniques firstly introduced by Corvino in
\cite{Cor00}. Perhaps one day numerical simulations will allow us to
contrast the physical differences (if any) of a spacetime evolved,
say, from Bowen-York initial data ---which shall not peel--- and those
of the time development of a data and initial data set which exactly is 
Bowen-York inside a compact set, and exactly Kerrian in the asymptotic
region ---which, arguably, should peel.

\medskip
The present article and a companion one ---which we shall refer as to
paper II--- extend the analysis initiated in
\cite{Fri98a,Val04a,Val04c} to the case of non-time symmetric (or time
asymmetric) initial data sets. In particular, the present article is
concerned with those time asymmetric initial data sets which are
conformally flat. Thus, it encompasses, but it is not limited to, some
of the initial data sets which are routinely used in the numeric
simulations of spacetimes with black holes like the Bowen-York and
Brand-Br\"ugmann \cite{BowYor80,BraBru97}. We shall consider second
fundamental forms with higher order multipole moments like the one
used in \cite{DaiLouTak02}. On the other hand, paper II shall deal
with time asymmetric initial data sets which are non-conformally flat,
like the one that can be derived from the Kerr metric written in
Boyer-Lindquist coordinates. Most of the ideas and methods to be
employed in paper I will be developed and discussed in the present
article.

A computer algebra based approach, similar to the one used in
\cite{Val04a,Val04c} will reveal the existence of a hierarchy of
obstructions to the smoothness of null infinity. In contrast to the
time symmetric case where the vanishing of the obstructions implied a
certain degree of regularity at both future and past null infinity,
the obstructions arising when contemplating non-time symmetric initial
data can be time asymmetric. That is, there is a hierarchy of
obstructions ensuring the smoothness of $\mathscr{I}^+$ and another one for
$\mathscr{I}^-$, and these do not imply each other in general. This novel
feature opens the possibility of the existence of spacetimes where the
two disconnected parts of null infinity have different degrees of
regularity. The existence of this kind of situations had been
speculated long time ago by Walker \& Will
\cite{WalWil79a,WalWil79b}\footnote{In these two articles, the authors
discussed by means of post-Newtonian methods the effects of past null
infinity of the incoming gravitational radiation produced by two
gravitating bodies coming from infinity, $i^-$, on approximately
hyperbolic orbits which then escape to infinity, $i^+$.}.

The analysis undertaken in \cite{GarPri00,Val04b,Val04c} has shown
that the only stationary spacetime that can be expected to admit
conformally flat slices is the Schwarzschild spacetime. Thus, based on
prior experience, one would expect that the new hierarchy of
obstructions that are obtained in this analysis should, if vanishing,
restrict the initial data set to be asymptotic Schwarzschildean to a
certain order. This is found to be the case only if both the
obstructions at $\mathscr{I}^+$ and $\mathscr{I}^-$ vanish. The Schwarzschild data
referred to in this context is not the standard time symmetric
$t=constant$ slice obtained from writing the Schwarzschild metric in
isotropic coordinates, but a non-time symmetric, conformally flat
slice having a second fundamental form decaying like
\begin{equation}
\widetilde{\chi}_{ab}=\O\left(\frac{1}{|y|^{-1}}\right).
\end{equation}
as $|y|\rightarrow \infty$ in some coordinates adapted to the
asymptotic region ---see below. Whether it is possible to have a
spacetime whose future null infinity is fully smooth, without having
to restrict the behaviour of the field at past null infinity is a
question which would require a substantial refinement of the computer
algebra methods used in this article, and which will be left for the
future.

\bigskip
This article follows, in as much as it possible, the notation and
nomenclature of \cite{Fri98a}, and in particular that of
\cite{Fri04}. It makes use of a number of results and techniques which
although they are available in the literature, they may not be, widely
known. Therefore, some time is spend in introducing the necessary
ideas and concepts. The reader is in any case remitted to the original
articles for full details.

\bigskip
The article is structured as follows: in section 2, a discussion of
the constraint equations in the light of our intended applications is
performed. In particular, some results due to Dain \& Friedrich,
guaranteeing the existence of the type of initial data under
consideration are recalled. In section 3, a discussion of the
description of the geometry of the initial hypersurface near spatial
infinity in terms of spatial spinors is given. In section 4, a
discussion of the solutions to the conformally flat momentum
constraint within the space spinor formalism is given. Section 5 does
likewise with the Hamiltonian constraint. Section 6 is concerned with
the particularities of the propagation equations implied by the
Conformal Einstein equations in the region of spacetime near null and
spatial infinity. Section 7 deals with some properties of the
asymptotic expansions near null and spatial infinity introduced by
Friedrich. Section 8 discusses the results of our computer algebra
calculations. Finally, in section 9 some concluding remarks are given. 

\bigskip 
In order to perform the calculations here described, a number of
assumptions have been made. In order to help the reader to keep track
of them, they have been clearly marked in the text and numbered from 1 to
5.

\subsection{General conventions}
In this work we shall be considering spacetimes
$(\widetilde{\mathcal{M}},\widetilde{g}_{\mu\nu})$ arising as the
development of some Cauchy initial data
$(\widetilde{\mathcal{S}},\widetilde{h}_{\alpha\beta},\widetilde{\chi}_{\alpha\beta})$.
Tilded quantities will refer to quantities in the \emph{physical}
spacetime, whereas untilded ones will denote generically quantities on
an \emph{unphysical} ---i.e. conformally rescaled--- spacetime.  The
indices $\mu$, $\nu$, $\lambda,\ldots$ (second half of the Greek
alphabet) are spacetime indices taking the values $0,\ldots,3$; while
$\alpha$, $\beta$, $\gamma,\dots$ are spatial ones with range
$1,\ldots,3$. The latin indices $a$, $b$, $c,\ldots$ will be used in
spatial expressions which are valid for a particular coordinate system
(usually a Cartesian normal one) and take the values $1,\ldots,3$. The
indices, $i$, $j$, $k,\ldots$ are spatial frame indices ranging
$1,\ldots,3$, while $A$, $B$, $C,\ldots$ will be spinorial indices
taking the values $0$, $1$. Because of the use of spinors, the
signature of $\widetilde{g}_{\mu\nu}$ will be taken to be $(+,-,-,-)$,
and the 3-dimensional metric $\widetilde{h}_{\alpha\beta}$ will be
negative definite. Under our conventions and labeling, the Einstein
constraint equations read
\begin{subequations}
\begin{eqnarray}
&& \rwt-\chiwt^2+\chiwt_{\alpha\beta}\chiwt^{\alpha\beta}=0, \label{Hamiltonian} \\
&& \Dwt^{\alpha} \chiwt_{\alpha\beta} -\Dwt_{\beta}\chiwt =0 \label{momentum},
\end{eqnarray}
\end{subequations}
where $\widetilde{D}$ denotes the connection associated with the
3-metric $\widetilde{h}_{\alpha\beta}$, $\widetilde{r}$ is the corresponding Ricci
scalar and we have written
$\chiwt=\chiwt^{\alpha}_{\phantom{\alpha}\alpha}$.

The initial data set
$(\widetilde{\mathcal{S}},\widetilde{h}_{\alpha\beta},\widetilde{\chi}_{\alpha\beta})$
is to be assumed \emph{asymptotically flat} in the sense that
there exists a compact subset of $\widetilde{\mathcal{S}}$ such that its
complement is the union of disjoint subsets $\widetilde{\mathcal{S}}_k$,
$k=1,2,\ldots, K$ of $\widetilde{\mathcal{S}}$ ---the asymptotically flat
ends--- each of which can be identified diffeomorphically with
$\{\Real^3\;|\phantom{x} |y|>r_0$\} where $r_0$ is some positive real
number, and $|y|=\sqrt{ (y^1)^2+(y^2)^2+(y^3)^3}$. We shall require
that the coordinates so introduced satisfy
\begin{equation} \label{decay_1}
\widetilde{h}_{ab}=-\left(1+\frac{2m_k}{|y|}\right)\delta_{ab}+\O\left(\frac{1}{|y|^2}\right), \quad \mbox{as } |y|\rightarrow \infty, \quad m_k=constant.
\end{equation}
Moreover, we shall require the second fundamental form to decay as
\begin{equation} \label{decay_2}
\chiwt_{ab}=\O\left(\frac{1}{|y|^2}\right),  \quad \mbox{as } |y|\rightarrow \infty.
\end{equation}
Besides the above asymptotic flatness requirements, we shall further
assume $(\widetilde{\mathcal{S}},\widetilde{h}_{\alpha\beta})$ to be
\emph{asymptotically Euclidean and regular} in the sense that there
exists a 3-dimensional, orientable, smooth, compact Riemannian
manifold $(\mathcal{S},h_{\alpha\beta})$ with points $i_k\in \mathcal{S}$, $k=1,2,\ldots
q$, $q$ some positive integer, a diffeomorphism $\Phi$ of
$\mathcal{S}\setminus\{i_1,\ldots,i_q\}$ onto $\widetilde{\mathcal{S}}$ and a function
$\Omega\in C^2(\mathcal{S})\cap C^\infty(\mathcal{S}\setminus\{i_1,\ldots,i_q\})$ with the
following properties:
\begin{subequations}
\begin{eqnarray}
&& \Omega=0, \quad d\Omega=0, \quad \mbox{Hess}(\Omega)_{\alpha\beta}=-2h_{\alpha\beta}, \label{decay_3}\\
&& \Omega>0 \mbox{ on } \mathcal{S}\setminus\{i_1,\ldots,i_q\}, \label{decay_4}\\
&& h_{\alpha\beta}=\Omega^2\Phi_*\widetilde{h}_{\alpha\beta} \mbox{ on } \mathcal{S}\setminus\{i_1,\ldots,i_q\}. \label{decay_5} 
\end{eqnarray}
\end{subequations}
Note that suitable punctured neighbourhoods of the points $i_k$
correspond to the asymptotically flat ends of
$(\widetilde{\mathcal{S}},\widetilde{h}_{\alpha\beta})$. Thus, each point $i_k$
represents a spatial infinity. The discussion in this article will be
concerned with the behaviour of the gravitational field in one of
these spatial infinities. So, without loss of generality we assume
that there is one of them, which we denote by $i$. The manifold $S$
will be identified via $\Phi$ with $\mathcal{S}\setminus\{i\}$.

\section{The conformal constraint equations}

In the sequel, we shall consider asymptotic expansions near infinity
for a wide class of solutions of the constraint equations
(\ref{Hamiltonian}) and (\ref{momentum}). Instead of working in the
physical spacetime manifold, we shall consider a conformally rescaled
version thereof
\begin{equation}
g_{\mu\nu}=\Omega^2\widetilde{g}_{\mu\nu}.
\end{equation}
The first and second fundamental forms induced by the metrics
$g_{\mu\nu}$ and $\widetilde{g}_{\mu\nu}$ on $\widetilde{\mathcal{S}}$ are then
related via
\begin{equation}
h_{\alpha\beta}=\Omega^2\widetilde{h}_{\alpha\beta}, \quad \chi_{\alpha\beta}=\Omega(\widetilde{\chi}_{\alpha\beta} + \Sigma \widetilde{h}_{\alpha\beta}),
\end{equation}
where $\Sigma$ denotes the derivative of $\Omega$ along the future
directed $g$-unit normal of $\widetilde{\mathcal{S}}$. As a consequence of the
latter two equations the traces
$\chi=h^{\alpha\beta}\chi_{\alpha\beta}$,
$\widetilde{\chi}=\widetilde{h}^{\alpha\beta}\widetilde{\chi}_{\alpha\beta}$
satisfy the relation
\begin{equation}
\Omega \chi =\widetilde{\chi}+3\Sigma.
\end{equation} 
The constraint equations (\ref{Hamiltonian}) and (\ref{momentum}) can
be written in terms of the conformal fields as ---see \cite{Fri88}---
\begin{subequations}
\begin{eqnarray}
&& 2\Omega D_{\alpha}D^\alpha\Omega -3 D_\alpha\Omega D^\alpha \Omega +\frac{1}{2}\Omega^2 r-3\Sigma^2 -\frac{1}{2}\Omega^2 -\frac{1}{2}\left( \chi^2-\chi_{\alpha\beta} \chi^{\alpha\beta} \right) + 2\Omega \Sigma \chi =0, \label{conformal_Hamiltonian}\\
&& \Omega^3 D^\alpha(\Omega^{-2} \chi_{\alpha\beta}) -\Omega\left( D_\beta \chi -2 \Omega^{-1} D_\beta \Sigma \right)=0, \label{conformal_momentum}
\end{eqnarray}
\end{subequations}
where $D_\alpha$ denotes the Levi-Civita connection and $r$ the
Ricci scalar of the metric $h_{\alpha\beta}$. The latter expressions lead to our first

\begin{assumption}[maximal initial data] \label{maximal}
In this work we shall assume that
\begin{equation}
\Sigma=0, \quad \chi=0 \mbox{ on } \widetilde{\mathcal{S}}.
\end{equation}
\end{assumption}
 
The first condition has to do with the choice of the conformal factor
$\Omega$, which up to now remains unspecified. Together they state
that our investigation will be restricted to hypersurfaces
$\widetilde{\mathcal{S}}$ which are maximal with respect to both metrics
$g_{\mu\nu}$ and $\widetilde{g}_{\mu\nu}$ so that we can make use of
the so-called conformal Ansatz to study the solutions of the
constraint equations. Under assumption \ref{maximal}, the conformal
constraint equations on $\widetilde{\mathcal{S}}$ reduce to
\begin{eqnarray}
&& \left(D_\alpha D^\alpha -\frac{1}{8}r\right) \vartheta =\frac{1}{8}\chi_{\alpha\beta}\chi^{\alpha\beta} \vartheta \mbox{ with } \vartheta=\Omega^{-1/2}, \label{maximal_Hamiltonian} \\
&& D^\alpha \left(\Omega^{-2}\chi_{\alpha\beta} \right)=0. \label{maximal_momentum}
\end{eqnarray}

In order to solve in terms of expansions the constraint equations we
shall make use of a slight variation of the so-called \emph{conformal
method} consisting of the following steps:

\begin{itemize}
\item[(i)] choose a negative definite metric $h_{\alpha\beta}$ on a 3-dimensional,
orientable, smooth, compact manifold $\mathcal{S}$. In $\mathcal{S}$, choose a point $i$
and set $\widetilde{\mathcal{S}}=\mathcal{S}\setminus{i}$.

\item[(ii)] Find a symmetric, $h$-tracefree tensor field
$\psi_{\alpha\beta}$ on $\widetilde{S}$ satisfying
\begin{equation}
D^\alpha \psi_{\alpha\beta}=0. \label{rescaledmomentum}
\end{equation}
A standard way of finding the tensor $\psi_{\alpha\beta}$ is by means
of a York splitting. That is, choose a symmetric, $h$-tracefree tensor
$\psi'_{\alpha\beta}$ on $\widetilde{S}$ and set
\begin{equation}
\psi_{\alpha\beta}= \psi'_{\alpha\beta} + D_{\alpha} v_\beta + D_\beta v_\alpha -\frac{2}{3}h_{\alpha\beta}D_\gamma v^\gamma, \label{york_splitting}
\end{equation}
where $v_\alpha$ is some 1-form. It is well known that the latter
Ansatz upon substitution on equation (\ref{maximal_momentum}) leads to
an elliptic equation for $v_\alpha$.

\item[(iii)] Finally, substitute
$\psi_{\alpha\beta}=\vartheta^{-4}\psi_{\alpha\beta}$ into equation
(\ref{maximal_Hamiltonian}) and find a positive solution $\vartheta$
of the resulting \emph{Licnerowicz equation}:
\begin{equation} \label{Licnerowicz}
\left(D_\alpha D^\alpha -\frac{1}{8}r\right)\vartheta =\frac{1}{8}\psi_{\alpha\beta}\psi^{\alpha\beta}\vartheta^{-7}.
\end{equation}
\end{itemize}
The fields $h_{\alpha\beta}$, $\Omega=\vartheta^{2}$ and
$\chi_{\alpha\beta}=\Omega^2\psi_{\alpha\beta}$ thus obtained provide
a solution to the conformal constraint equations
(\ref{maximal_Hamiltonian}) and (\ref{maximal_momentum}).

\bigskip
The above procedure is to be supplemented with asymptotic (boundary)
conditions for the fields $\vartheta$ and $\psi_{\alpha\beta}$
consistent with the conditions (\ref{decay_1}) and
(\ref{decay_2}). Consider $a>0$ small enough so that the $h-$metric
ball $\mathcal{B}_a(i)$ centred on $i$ is a strictly convex normal neighbourhood of
$i$, and let $\{x^a\}$ be normal coordinates with origin at $i$ based
on an $h-$orthonormal frame $\{e_j\}$ at $i$. Consistent with the
conditions (\ref{decay_1})-(\ref{decay_2}) and
(\ref{decay_3})-(\ref{decay_5}) we shall require that
\begin{equation}\label{boundary_1}
|x|\vartheta \rightarrow 1, \quad \mbox{ as } |x|\rightarrow 0.
\end{equation}
The tensorial fields $\widetilde{\chi}_{\alpha\beta}$, $\chi_{\alpha\beta}$, $\psi_{\alpha\beta}$ are related by
\begin{equation}
\widetilde{\chi}_{\alpha\beta}=\Omega^{-1}\chi_{\alpha\beta}=\Omega\psi_{\alpha\beta}
\end{equation} 
Thus, the appropriate behaviour for the unphysical fields is given by
\begin{equation} \label{boundary_2}
\chi_{\alpha\beta}=\O(1), \quad \psi_{\alpha\beta}=\O\left(\frac{1}{|x|^4}\right) \quad \mbox{ as } |x|\rightarrow 0.
\end{equation}

\bigskip
In contrast to the analysis carried out in \cite{Fri98a}, here we
shall consider initial data sets ---i.e. solutions to the conformal
constraint equations--- which are not necessarily time symmetric, that
is, in general we shall assume that $\chi_{\alpha\beta}\neq
0$. Discussions given in \cite{Fri98a,Val04a,Val04b,Val04c} will arise
as particular cases of our present analysis. This first article shall
be concerned with the analysis of conformally flat initial data
sets. Accordingly, we make the following assumption

\begin{assumption}[conformal flatness]\label{cf}
It will beassumed that there is a neighbourhood, $\mathcal{B}_a(i) \subset
\mathcal{S}$ of radius $a$ centred on $i$ and coordinates $\{x^a\}$
for which the (unphysical) conformal metric is of the form
\begin{equation}
h_{ab}=-\delta_{ab}=\mbox{diag}(-1,-1,-1).
\end{equation} 
\end{assumption}

\bigskip
Results regarding the nonexistence of conformally flat slices
imply that the latter assumption eliminates all stationary initial
data sets except (time asymmetric) Schwarzschildean ones from our
considerations \cite{GarPri00,Val04b,Val04c} ---this observation will
play an important role in the interpretation of the results of our
investigations. Moreover, based on the results of Dain
\cite{Dai01b,Dai01c} it seems that any discussion of strictly
stationary data ---i.e. data whose development is stationary but not
static--- has to consider conformal metrics which are non-smooth. For
example, the conformal metric one obtains for Kerrian initial data
form $t=constant$ slices in Boyer-Lindquist coordinates is just
$C^{2,\alpha}$ at infinity\footnote{$f\in C^{p,\alpha}$ means that the
function has $p$-th order derivatives f which are H\"older continuous
with exponent $\alpha$. A function $f$ is said to be H\"older
continuous with exponent $\alpha$ at a point $x_0$ if there is a
constant $C$ such that $|f(x)-f(x_0)|\leq C|x-x_0|^\alpha$,
$0<\alpha<1$ for $x$ in a neighbourhood of $x_0$.}. This
non-smoothness of the conformal metric shall be discussed \emph{in extensis} in
paper II.

\subsection{Existence of solutions to the conformally flat constraint equations}
Under the assumptions (\ref{maximal}) and (\ref{cf}) the Einstein
constraint equations reduce ---in Cartesian coordinates--- to
\begin{subequations}
\begin{eqnarray}
&& \partial_a \psi^{ab}=0, \\
&& \Delta\vartheta =\frac{1}{8}\psi_{ab}\psi^{ab}\vartheta^{-7},
\end{eqnarray}
\end{subequations}
together with the boundary conditions (\ref{boundary_1}) and
(\ref{boundary_2}).  We want to consider solutions to these equations
which can be expanded in powers of $|x|$ ---note that $f(x^a)=|x|$ is not a smooth function of the coordinates $x^a$ as $\partial_a f=-x^a/|x|$. In particular, we shall
exclude from our discussion any solutions whose expansions contain
fractional powers of $|x|$ or $\ln |x|$ terms. These considerations lead
naturally to the following definition, which we retake from
\cite{DaiFri01}:

\begin{definition}[$E^\infty spaces$]
A function $f\in C^\infty(\widetilde{\mathcal{S}})$ is said to be in
$E^\infty(\mathcal{B}_a(i))$ if on $\mathcal{B}_a(i)$ one can write $f=f_1 + |x|f_2$ with
$f_1, \;f_2\in C^\infty(\mathcal{B}_a(i))$.
\end{definition}

The type of solutions of the constraint equations that we want to
consider are naturally described in terms of these spaces. The
existence of the solutions to (\ref{Licnerowicz}) and
(\ref{rescaledmomentum}) are guaranteed by the following two results
due to Dain \& Friedrich \cite{DaiFri01}. Note that these results
apply to a larger class of data than the one under consideration
---here $h_{ab}=\delta_{ab}$, which is trivially smooth. However, these are not
general enough to encompass strictly stationary initial data.

\begin{theorem}[Dain \& Friedrich, 2001] \label{existence Hamiltonian}
Let $h_{ab}$ be a smooth metric on $\mathcal{S}$ with non-negative Ricci
scalar $r$. Assume that $\widetilde{\psi}_{ab}$ is smooth on
$\widetilde{\mathcal{S}}$ and satisfies on $\mathcal{B}_a(i)$ 
\begin{equation}
|x|^8\psi_{ab}\psi^{ab}\in E^\infty(\mathcal{B}_a(i)). 
\end{equation}
Then there exists a unique solution of equation (\ref{Licnerowicz})
which is positive, satisfies the boundary conditions
(\ref{boundary_1}) and (\ref{boundary_2}) and in $\mathcal{B}_a(i)$ has the form
\begin{equation}
\vartheta =\frac{\hat{\vartheta}}{|x|}, \quad \hat{\vartheta}\in E^\infty(\mathcal{B}_a(i)), \quad \hat{\vartheta}(i)=1.
\end{equation}
\end{theorem}

The condition $|x|^8\psi_{ab}\psi^{ab}\in E^\infty(\mathcal{B}_a(i))$
excludes the possibility of second fundamental forms with linear
momentum ---the so-called boosted slices. In \cite{DaiFri01} it has
been shown that if one considers a smooth conformal metric and second
fundamental forms with linear momentum, then the solution of the
Licnerowicz equation will contain logarithmic terms at order $|x|^2$.

Theorem 1 is complemented by the following result stating the
conditions for the existence of a second fundamental form of the
required type. It reads

\begin{theorem}[Dain \& Friedrich, 2001] \label{existence momentum}
Let $h_{ab}$ be a smooth metric on $\mathcal{S}$. There exist traceless
tensor fields $\psi_{ab}\in C^\infty(\mathcal{S}\setminus\{i\})$
satisfying equation (\ref{rescaledmomentum}) with the properties:
\begin{itemize}
\item[(i)] in the normal coordinates $\{x^a\}$,
\begin{equation}
\psi_{ab}=\frac{A}{|x|^3}(3n_a n_b -\delta_{ab}) + \frac{3}{|x|^3}(n_b \epsilon_{c a d}J^d n^c + n_a \epsilon_{d b c}J^c n^d) +\O(|x|^{-2}),
\end{equation}
with $n^a=x^a/|x|$.

\item[(ii)] $D^a \psi_{ab}=0$ on $\widetilde{\mathcal{S}}$

\item[(iii)] $|x|^8\psi_{ab}\psi^{ab}\in E^\infty(\mathcal{B}_a(i))$.

\end{itemize}  
\end{theorem}

The form of the leading terms of the extrinsic curvature given in the
latter theorem is, except for the absence of a term containing
linear momentum, that of the Bowen-York Ansatz, which is routinely used
in numerical simulations of black hole collisions ---see
e.g. \cite{BowYor80,BraBru97}. Note that the theorem allows for the
presence of higher order terms ---which, as it shall be shown in the
sequel, will encode the multipolar content of the second fundamental
form. These terms will play a crucial role in paper II when discussing
the structure of stationary data. In relation to this point, we note
that the results of \cite{Val04c} strongly suggest that the
only stationary data contained in the hypothesis of theorems 1 and 2
are non-time symmetric Schwarzschildean data. We shall return to this
point later.

\section{The initial slice near $i$}
Consider the Cartesian coordinates $\{x^a\}$ discussed in the previous
section, and the orthonormal frame $\{e_j\}$ such that
$e_3^a=x^a/|x|$. Under the assumptions made in the previous section
one has that the solutions of the constraint equations
(\ref{rescaledmomentum}) and (\ref{Licnerowicz}) lead to a conformal
factor of the form
\begin{equation}
\label{expansion_Omega}
\Omega=|x|^2-m|x|^3+\O(|x|)^4.
\end{equation}
Let $d^i_{\phantom{i}jkl}=\Omega^{-1}C^i_{\phantom{i}jkl}$ be the
\emph{rescaled Weyl tensor}, where $C^i_{\phantom{i}jkl}$ is the Weyl
tensor of the metric $g_{\mu\nu}$. The expansion
(\ref{expansion_Omega}) implies for the components of
$d^i_{\phantom{i}jkl}$ an expansion in $\mathcal{B}_a(i)$ of the form:
\begin{equation}
\label{Weyl divergent}
d_{ijkl}=\frac{m(h_{ik}-3\delta^3_i \delta^3_k)\delta^0_j\delta^0_l}{|x|^3}+\O\left(\frac{1}{|x|^2}\right),
\end{equation}
where $h_{ij}\equiv h(e_i,e_j)=h_{ab}e^a_i e^b_j=-\delta_{ij}$. In order to discuss
this kind of singular quantities, it is convenient to consider a
certain submanifold of the bundle of frames. We present here the crucial points of the construction of this manifold. For full details the reader is remitted to  \cite{Fri98a}, \cite{Fri04} or \cite{Fri03a}.

\subsection{The manifold $\mathcal{C}_a$}
In what follows we shall be using a space spinor formalism analogous
to a tensorial 3+1 decomposition\footnote{However, it must be said
that the use of such a spinorial formalism is not essential. An
equivalent discussion can be carried out in terms of
frames. See e.g. \cite{Fri03a}}. Consider the (unphysical) spacetime $(\mathcal{M},g_{\mu\nu})$
obtained as the development of the initial data set
$(\mathcal{S},h_{\alpha\beta},\chi_{\alpha\beta})$. Let $SL(\mathcal{S})$ be the set of
spin dyads $\delta=\{\delta_A\}_{A=0,1}$ on $\mathcal{S}$ which are normalised
with respect to the alternating form $\epsilon$ in such a way that
\begin{equation}
\epsilon(\delta_A,\delta_B)=\epsilon_{AB}, \quad \epsilon_{01}=1.
\end{equation}
The set $SL(\mathcal{S})$ has a natural bundle structure where
$\mathcal{S}$ is the base space,  and its structure group is given by
\begin{equation}
SL(2,\Complex)=\{t^A_{\phantom{A}B}\in GL(2,\Complex)\;|\; \epsilon_{AC}t^A_{\phantom{A}B} t^C_{\phantom{C}D}=\epsilon_{BD}\},
\end{equation}
acting on $SL(\mathcal{S})$ by $\delta\mapsto \delta\cdot t=\{\delta_A t^A_{\phantom{A}B}\}_{B=0,1}$. Now, let $\tau=\sqrt{2}e_0$, where $e_0$ is the future $g$-unit normal of $\mathcal{S}$ and
\begin{equation}
\tau_{AA'}=g(\tau,\delta_A\overline{\delta}_{A'})=\epsilon_{A}^{\phantom{A}0}\epsilon_{A'}^{\phantom{A'}0'}+\epsilon_{A}^{\phantom{A}1}\epsilon_{A'}^{\phantom{A'}1'}
\end{equation}
is its spinorial counterpart --- that is, $\tau=\tau^i e_i =
\sigma^i_{AA'}\tau^{AA'}e_i$ where $\sigma^i_{AA'}$ denote the
Infeld-van der Waerden symbols and $\{e_i\}_{i=0,\ldots,3}$ is an
orthonormal frame. The spinor $\tau_{AA'}$ enables the introduction
of space-spinors ---sometimes also called $SU(2)$ spinors, see
\cite{Ash91,Fra98a,Som80}: it defines a subbundle $SU(\mathcal{S})$ of $SL(\mathcal{S})$
with structure group
\begin{equation}
SU(2,\Complex)=\{ t^A_{\phantom{A}B}\in SL(2,\Complex)\; |\; \tau_{AA'}t^A_{\phantom{A}B}\overline{t}^{A'}_{\phantom{A'}B'}=\tau_{BB'}\},
\end{equation}   
and spatial van der Waerden symbols:
\begin{equation}
\sigma_{i}^{AB}=\sigma^{(A}_{i\phantom{(A}A'}\tau^{B)A'}, \quad \sigma^i_{AB}=\tau_{(B}^{\phantom{(B}A'}\sigma^{i}_{\phantom{i}A)A'}, \quad i=1,2,3.
\end{equation}
The latter satisfy
\begin{equation}
h_{ij}=\sigma_{iAB}\sigma_j^{AB}, \quad -\delta_{ij}\sigma^i_{AB}\sigma^j_{CD}=-\epsilon_{A(C}\epsilon_{D)B}\equiv h_{ABCD},
\end{equation}
with $h_{ij}=h(e_i,e_j)=-\delta_{ij}$. The bundle $SU(\mathcal{S})$ can be
endowed with a $\mathfrak{su}(2,\Complex)$-valued \emph{connection form}
$\check{\omega}^A_{\phantom{A}B}$ compatible with the metric
$h_{\alpha\beta}$ and 1-form $\sigma^{AB}$, the \emph{solder form} of
$SU(S)$\footnote{To be more precise, these structures are inherited
from their analogues on $O_+(S)$, the bundle of positively oriented
orthonormal frames via the covering map of $SU(2)$ onto the
connected component $SO(3)$ of the rotation group given by $ SU(2) \ni
t^A_{\phantom{A}B} \stackrel{\Psi}{\longrightarrow}
t^{i}_{\phantom{i}j} =\sigma^i_{AB} t^A_{\phantom{A}C}
t^B_{\phantom{B}D} \sigma^{CD}_j \in SO(3)$, the induced isomorphism
of Lie algebras , $\Psi_*$ and suitable contractions with the
Infeld-van der Waerden symbols.}. The solder form satisfies by
construction
\begin{equation}
\label{metric}
h\equiv h_{\alpha\beta} dx^\alpha \otimes dx^\beta = h_{ABCD} \sigma^{AB}\otimes \sigma^{CD},
\end{equation}
where $\sigma^{AB}=\sigma^{AB}_{\alpha} dx^\alpha$ ---note that the $\sigma^{AB}_\alpha$ are not the spatial Infeld-van der Waerden symbols! 

Now, given a spinorial dyad $\delta\in SU(\mathcal{S})$ one can define an
 associated vector frame on $O_+(\mathcal{S})$ via
 $e_j=e_j(\delta)=\sigma^{AB}_j
 \delta_A\tau_B^{\phantom{B}B'}\overline{\delta}_{B'}$. It is noted
 that different dyads $\delta$ and $\delta^\prime$ can give rise to
 the same frame as long as they are $U(1)$-related\footnote{ Here we
 shall use the realisation
\[
U(1)=\left\{ t\in U(1)\phantom{X} |\phantom{X} t=\left(
\begin{array}{cc}
e^{i\phi} & 0 \\
0         & e^{-i\phi}
\end{array}\right), 
\phantom{X} \phi \in \Real \right\}.
\] 
Thus, $\delta$ and $\delta^\prime$ are said to be $U(1)$-related if
$\delta^\prime_A= t^B_{\phantom{B}A}\delta_B$ with
$t=(t^A_{\phantom{A}B})\in U(1)$.}. We shall restrict our attention to
dyads giving rise to frames $\{e_j\}_{j=0,\cdots,3}$ on $\mathcal{B}_a(i)$ such
that $e_3$ is tangent to the $h$-geodesics starting at $i$. Let
$\check{H}$ denote the horizontal vector field on $SU(\mathcal{S})$
projecting to the aforediscussed radial vectors $e_3$.

\medskip
The fiber $\pi^{-1}(i)\subset SU(\mathcal{S})$ ---the fiber ``over'' $i$--- can be
parametrised by choosing a fixed dyad $\delta^*$ and then letting the
group $SU(2,\Complex)$ act on it. Let $(-a,a)\ni \rho \mapsto
\delta(\rho,t)\in SU(\mathcal{S})$ be the integral curve to the vector
$\check{H}$ satisfying $\delta(0,t)=\delta(t)\in \pi^{-1}(i)$. With
this notation we define the set
\begin{equation}
\mathcal{C}_a=\bigg \{ \delta(\rho,t)\in SU(\mathcal{S}) \phantom{X}|\phantom{X}  |\rho|<a, \phantom{X} t\in SU(2,\Complex) \bigg \}
\end{equation}
which is a smooth submanifold of $SU(\mathcal{S})$ diffeomorphic to
$(-a,a)\times SU(2,\Complex)$. The vector field $\check{H}$ is such
that its integral curves through the fiber $\pi^{-1}(i)$ project onto
the geodesics through $i$. From here it follows that the
projection map $\pi$ of the bundle $SU(\mathcal{S})$ maps $\mathcal{C}_a$ into $\mathcal{B}_a(i)$.

Let in the sequel $\mathcal{I}^0\equiv \pi^{-1}(i)=\{\rho=0\}$ denote
the fiber over $i$. It can be seen that $\mathcal{I}^0\approx
SU(2,\Complex)$. On the other hand, for $p\in
\mathcal{B}_a(i)\setminus \{i\}$ it turns out that $\pi^{-1}(p)$
consists of an orbit of $U(1)$ for which $\rho=|x(p)|$, and another
for which $\rho=-|x(p)|$, where $x^a(p)$ denote normal coordinates of
the point $p$. Because of the latter, it is convenient, in order to
understand better the structure of the manifold $\mathcal{C}_a$ to
quotient out the effect of $U(1)$ on $\mathcal{C}_a(i)$. The projection
map restricted to $\mathcal{C}_a\subset SU(\mathcal{S})$ can be
factorised as $\mathcal{C}_a \stackrel{\pi_1}{\longrightarrow}
\mathcal{C}^\prime_a=\mathcal{C}_a/U(1)
\stackrel{\pi_2}{\longrightarrow} \mathcal{B}_a(i)$. Now, the set
$\pi^{-1}_2(\mathcal{B}_a(i))$ consists of a component
$\mathcal{C}^{\prime+}_a$ on which $\rho>0$ and another,
$\mathcal{C}^{\prime-}_a$ on which $\rho<0$. The projection $\pi_2$
maps both $\mathcal{C}^{\prime+}_a$ and $\mathcal{C}^{\prime-}_a$ into
the punctured disk $\mathcal{B}_a(i)\setminus\{i\}$. This last fact
can be used to identify $\mathcal{B}_a(i)$ with
$\mathcal{C}^{\prime+}_a$ acquiring in the process a boundary given by
$\pi_1(\mathcal{I}^0)=\pi^{-1}_2(i)\approx \mathcal{S}^2$. The set
$\overline{\mathcal{B}}_a(i)=\left(\mathcal{B}_a(i)\setminus\{i\}\right)\cup\pi^{-1}_2(i)\approx[0,a)\times
\mathcal{S}^2$ is a smooth manifold with boundary. Summarising, we
have obtained an extension of the initial (physical) manifold
$\widetilde{\mathcal{S}}$ by blowing up the point the point $i$ into
a sphere.

For practical purposes, it is more convenient to make use of the
4-dimensional $U(1)$-bundle
\begin{equation}
\overline{\mathcal{C}}^+_a=\mathcal{C}^+_a \cup \mathcal{I}^0 =\bigg \{\delta=\delta(\rho,t)\in \mathcal{C}_a\phantom{X} | \phantom{X} \rho(\delta) \geq 0 \bigg \} \approx [0,a)\times SU(2,\Complex) .
\end{equation}
The construction of the manifold $\mathcal{C}_a$ is such that a
number of useful structures are inherited by $\mathcal{C}_a$ from $SU(\mathcal{S})$. In
particular, the solder and connection forms can be pulled back to
smooth 1-forms on $\mathcal{C}_a$. These shall be again denoted by $\sigma^{AB}$ and $\check{\omega}^{A}_{\phantom{A}B}$. They satisfy the \emph{structure equations}:
\begin{subequations}
\begin{eqnarray}
&& d\sigma^{AB} = -\check{\omega}^A_{\phantom{A}E}\wedge
\sigma^{EB}-\check{\omega}^B_{\phantom{B}E}\wedge \sigma^{AE}, \\ &&
d\check{\omega}^A_{\phantom{A}B}=-\check{\omega}^A_{\phantom{A}E}\wedge
\check{\omega}^E_{\phantom{E}B} + \check{\Omega}^{A}_{\phantom{A}B},
\end{eqnarray}
\end{subequations}
with
\begin{equation}
\check{\Omega}^A_{\phantom{A}B}=\frac{1}{2} r^{A}_{\phantom{A}BCDEF} \sigma^{CD}\wedge \sigma^{EF},
\end{equation}
the so-called \emph{curvature form} determined by the \emph{curvature spinor} $r_{ABCDEF}$ given by
\begin{equation}
r_{ABCDEF}=\bigg( \frac{1}{2}s_{ABCE}-\frac{1}{12}r h_{ABCE}\bigg) \epsilon_{DF} +\bigg( \frac{1}{2} s_{ABDF}-\frac{1}{12}r h_{ABDF} \bigg)\epsilon_{CE},
\end{equation}
where $s_{ABCD}=s_{(ABCD)}$ is the tracefree part of the Ricci tensor
of $h_{\alpha\beta}$ and $r$ its Ricci scalar. These satisfy the \emph{3-dimensional Bianchi identity}
\begin{equation}
D^{AB}s_{ABCD}=\frac{1}{6}D_{CD}r.
\end{equation}

In the case of the data considered in this paper ---conformally flat,
assumption 2--- it follows that $s_{ABCD}=0$ and $r=0$ so that
$\check{\Omega}^A_{\phantom{A}B}=0$. A detailed analysis of the
solutions of the structure equations in the case of non-smooth
conformal metrics will be given in paper II.

\subsection{Lifts to $\mathcal{C}_a$ and reality conditions}
One can use the function $\rho$ and the matrices
$t=(t^A_{\phantom{A}B})\in SU(2,\Complex)$ to coordinatise
$\mathcal{C}_a$. Again, let $\{x^a\}$ denote normal coordinates on
$\mathcal{B}_a(i)$ such that $x^a(i)=0$. The projection map $\pi$ of
$\mathcal{C}_a$ onto $\mathcal{B}_a(i)$ has the local expression
\begin{equation}
\pi:(\rho,t)\mapsto x^a(\rho,t)=\sqrt{2}\rho\delta^a_j\sigma^j_{CD}t^C_{\phantom{C}0}t^D_{\phantom{D}1}.
\end{equation}
This last expression can be used to ``lift'' scalar and spinorial
fields on $\mathcal{B}_a(i)$ to $\mathcal{C}_a$. In an abuse of notation we shall,
generally, denote the lift to $\mathcal{C}_a$ of a spinorial field
$\mu_{AB\cdots EF}$ on $\mathcal{B}_a(i)$ again by $\mu_{AB\cdots EF}$. To which
manifold the field belongs will be clear from the context.

The following two maps between spinors will play a role in the sequel:
 given a ``primed'' spinorial field $\mu_{A'}$, one can associate an
 unprimed one via $\mu_{A'}\mapsto
 \mu_A=\tau_{A}^{\phantom{A}A'}\mu_{A'}$. The other one is the \emph{Hermitian conjugation
 map} $\mu_A\mapsto
 \mu^+_A=\tau_A^{\phantom{A}A'}\overline{\mu}_{A'}$. These maps are
 extended to higher valence spinors in a direct way. A spinorial field
 $\nu_{A_1A'_1\cdots A_kA'_k}$ is said to be \emph{spatial} if and
 only if $\tau^{A_jA'_j}\nu_{A_1A_1'\cdots A_jA'_j \cdots A_kA'_k}=0$
 for $j=1,\ldots,k$. In which case its ``unprimed'' version is
 given by
\begin{equation}
\mu_{A_1B_1\cdots A_kB_k}=\tau_{B_1}^{\phantom{B_1}A'_1}\cdots \tau_{B_k}^{\phantom{B_k}A'_k}\nu_{A_1A'_1\cdots A_kA'_k}=\nu_{(A_1B_1)\cdots(A_kB_k)}.
\end{equation}
Conversely, any spinor field with $2k$ unprimed indices corresponds to
a spatial spinor. Thus, for a given spinorial field $\mu_{AA'}$ one
obtains an \emph{orthogonal splitting} of the form
\begin{equation}
\mu_{AA'}=\tau_{CA'}\tau^{CB'}\mu_{AB'}=\frac{1}{2}\tau_{AA'}\tau^{CC'}\mu_{CC'}-\tau^C_{\phantom{C}A'}\mu_{CA},
\end{equation}
where $\mu_{CA}=\tau_{(C}^{\phantom{(C}B'}\mu_{A)B'}$ corresponds to
the spatial part of $\mu_{AA'}$. Again, this procedure can be extended
in a standard way to higher valence spinors. As a final remark we note
that, given a spatial spinor field $\mu_{A_1B_1\cdots
A_kB_k}=\mu_{(A_1B_1)\cdots (A_kB_k)}$ the frame components
$\mu_{j_1\cdots
j_k}=\sigma^{A_1B_1}_{j_1}\cdots\sigma^{A_kB_k}_{j_k}\mu_{A_1B_1\cdots
A_kB_k}$ correspond to those of a real tensor field if and only if it
satisfies the reality condition
\begin{equation} \label{reality condition}
\mu^+_{A_1B_1\cdots A_k B_k}=(-1)^k\mu_{A_1B_1\cdots A_kB_k}.
\end{equation}

\subsection{Decompositions in terms of irreducible spinors}
A symmetric valence 2 spinor $\mu_{AB}$ has 3 essential components. In order to make this explicit, we make use of the irreducible spinors
\begin{equation}
x_{AB} \equiv \sqrt{2}\epsilon_{(A}^{\phantom{(A}0} \epsilon_{B)}^{\phantom{B}1}, \quad y_{AB}\equiv -\frac{1}{\sqrt{2}}\epsilon_A^{\phantom{A}1}\epsilon_B^{\phantom{B}1}, \quad z_{AB}\equiv \frac{1}{\sqrt{2}}\epsilon_A^{\phantom{A}0}\epsilon_B^{\phantom{B}0},  
\end{equation}
to write
\begin{equation}
\label{rank_2}
\mu_{AB}=\mu_x x_{AB} +\mu_y y_{AB} +\mu_z z_{AB}.
\end{equation}
Similarly, a valence 4 spinor $\nu_{ABCD}=\nu_{(AB)(CD)}$ has 9 essential components and can be written as
\begin{eqnarray}
&&\hspace{-1.5cm}\nu_{ABCD}=\nu_0\epsilon^0_{ABCD}+\nu_1\epsilon^1_{ABCD}+\nu_2\epsilon^2_{ABCD}+\nu_3\epsilon^3_{ABCD}+\nu_4\epsilon^4_{ABCD} \nonumber\\
&&\hspace{-1.2cm} +\nu_x (\epsilon_{AC}x_{BD}+\epsilon_{BD}x_{AC}) +\nu_y (\epsilon_{AC}y_{BD}+\epsilon_{BD}y_{AC}) + \nu_z(\epsilon_{AC}z_{BD}+\epsilon_{BD}z_{AC}) + \nu_h h_{ABCD}, \label{rank_4}
\end{eqnarray}
where
$\epsilon^i_{ABCD}=\epsilon_{(A}^{\phantom{(A}(E}\epsilon_B^{\phantom{B}F}\epsilon_C^{\phantom{C}G}\epsilon_{D)}^{\phantom{D)}H)_i}$. The
notation, which shall be used again in the sequel, $(\cdots)_i$ means
that the indices are to be symmetrised and then $i$ of them should be
set equal to $1$. In particular, if $\nu_{ABCD}$ is associated with a
symmetric tensor $\nu_{\alpha\beta}=\nu_{(\alpha\beta)}$ then
$\nu_x=\nu_y=\nu_z=0$. If, furthermore, $\nu_{\alpha\beta}$ is
traceless then $\nu_h=0$.

\subsection{Vector fields on $\mathcal{C}_a$}
Consequently with the discussion of the previous section one has that
$\check{H}=\partial_\rho$. Vector fields relative to the
$SU(2,\Complex)$-dependent part of the coordinates are obtained by
looking at the basis of the (3-dimensional) Lie algebra
$\mathfrak{su}(2,\Complex)$ given by
\begin{equation}
u_1=\frac{1}{2}\left(
\begin{array}{cc}
0 & i \\
i & 0
\end{array}\right), \quad
u_2=\frac{1}{2}\left(
\begin{array}{cc}
0 & -1 \\
1 & 0
\end{array}\right), \quad
u_3=\frac{1}{2}\left(
\begin{array}{cc}
i & 0 \\
0 & -i
\end{array}\right).
\end{equation}  
In particular, the vector $u_3$ is the generator of $U(1)$. Denote by
$Z_i$, $i=1,2,3$ the Killing vectors generated on $SU(\mathcal{S})$ by
$u_i$ and the action of $SU(2,\Complex)$. The vectors $Z_i$ are
tangent to $\mathcal{I}^0$. On $\mathcal{I}^0$ we set
\begin{equation}
X_+=-(Z_2+iZ_1), \quad X_-=-(Z_2-iZ_1), \quad X=-2iZ_3,
\end{equation}
and extend these vector fields to the rest of $\mathcal{C}_a$ by demanding them
to commute with $\check{H}=\partial_\rho$. For latter use we note that
\begin{equation}
[X,X_+]=2X_+, \quad [X,X_-]=-2X_-, \quad [X_+,X_-]=-X.
\end{equation}
The vector fields are complex conjugates of each other in the sense
that for a given real-valued function $W$,
$\overline{X_-W}=X_+W$. More importantly, it can be seen that for
$p\in \mathcal{B}_a(i)\setminus\{i\}$ the projections of the fields $\check{H}$,
$X_\pm$ span the tangent space at $p$.

We define a frame $c_{AB}=c_{(AB)}$ dual to the solder forms
$\sigma^{CD}$ and require it not to pick components along the fibres
---i.e. along the direction of $X$. These requirements imply
\begin{equation}
\langle \sigma^{AB}, c_{CD} \rangle= h^{AB}_{\phantom{AB}CD}, \quad c_{CD}=c^1_{CD}\partial_\rho + c^+_{CD}X_+ + c^-_{CD}X_-.
\end{equation}
Let $\alpha^\pm$ and $\alpha$ be 1-forms on $\mathcal{C}_a$
annihilating the vector fields $\partial_\tau$, $\partial_\rho$ and
having with $X_\pm$ the non-vanishing pairings
\begin{equation}
\langle \alpha^+,X_+ \rangle = \langle \alpha^-, X_-\rangle = \langle \alpha, X \rangle =1.
\end{equation}
From the properties of the solder form $\sigma^{AB}$ one finds that
\begin{equation}
c^1_{AB}=x_{AB}, \quad c^+_{AB}=\frac{1}{\rho}z_{AB}+\check{c}^+_{AB}, \quad c^-_{AB}=\frac{1}{\rho}y_{AB}+\check{c}^-_{AB}.
\end{equation}
In particular, for the case under consideration ---assumption 2---, one has $\check{c}^\pm_{AB}=0$. Furthermore, one has that
\begin{equation}
\sigma^{AB}=-\frac{1}{\rho}x^{AB} d\rho -2 y^{AB}\alpha^+ - 2z^{AB}\alpha^-,
\end{equation}
so that the projection of (\ref{metric}) onto $\mathcal{B}_a(i)$ renders
\begin{equation}
h=-d\rho\otimes d\rho-2\rho^2(\alpha^+\otimes \alpha^- +\alpha^-\otimes \alpha^+)\equiv -d\rho^2-\rho^2d\sigma^2,
\end{equation} 
so that $\d\sigma^2=2(\alpha^+\otimes \alpha^- +\alpha^-\otimes \alpha^+)$ corresponds to the pull back of the standard metric on $\mathcal{S}^2$.

\subsection{The connection coefficients}
The connection coefficients are defined by contracting the connection form $\check{\omega}^A_{\phantom{A}B}$ with the frame $c_{AB}$. In general, we write
\begin{equation}
\gamma_{CD\phantom{A}B}^{\phantom{CD}A} \equiv \langle \check{\omega}^A_{\phantom{A}B},c_{CD} \rangle = \frac{1}{\rho} \gamma_{CD\phantom{A}B}^{*\phantom{D}A}+ \check{\gamma}_{CD\phantom{A}B}^{\phantom{CD}A},
\end{equation}
where,
\begin{equation}
\gamma^*_{ABCD}=\frac{1}{2}(\epsilon_{AC}x_{BD}+\epsilon_{BD}x_{AC}),
\end{equation}
denotes the singular part of the connection coefficients. The regular
part of the connection can be related to the frame coefficients
$c_{AB}$ via commutator equations which under assumption 2 are
trivially satisfied as $\check{\gamma}_{ABCD}=0$. We defer any further
discussion of these matters to paper II. 

Let $f$ be a smooth function on $\mathcal{B}_a(i)$. We denote again by $f$ its lift. The covariant derivative of $f$ is then given on $\mathcal{C}_a$ by
\begin{equation}
D_{AB}f=c_{AB}(f).
\end{equation} 
Similarly, let $\mu_{AB}$ represent both a smooth spinor field on $\mathcal{B}_a(i)$ and its lift to $\mathcal{C}_a$. Then the covariant derivative of $\mu_{AB}$ is given by
\begin{equation}
D_{AB}\mu_{CD}= c_{AB}(\mu_{CD})- \gamma_{AB\phantom{E}C}^{\phantom{AB}E}\mu_{ED}-\gamma_{AB\phantom{E}D}^{\phantom{AB}E}\mu_{CE}.
\end{equation}
Analoguos formulae hold for higher valence spinors.

\subsection{Normal expansions}

In order to study the behaviour of diverse fields on $\mathcal{C}_a$
near $\mathcal{I}^0$ in a detailed manner, we shall make use of a
certain type of expansions which are obtained by lifting the Taylor
expansions (along the radial direction) of the fields and then
conveniently symmetrising.

A notion which will be of much help is the following: a smooth
function $f$ on $\mathcal{C}_a$ is said to have \emph{spin weight s} if
\begin{equation}
Xf=2sf
\end{equation}
where $2s$ is an integer. If furthermore, $f$ is analytic in $\mathcal{B}_a(i)$
then on $\mathcal{C}_a$ it admits an expansion of the form
\begin{equation} 
f=\sum^\infty_{p=|s|} \frac{1}{p!} \sum_{q\in Q_p} \sum^{2q}_{m=0} f_{p;2q,m} \TT{2q}{m}{q-s} \rho^p, \label{normal expansion}
\end{equation}
where
\begin{equation}
Q_p=\bigg \{q\in \Integer^+\cup\{0\} \phantom{X}|\phantom{X} |s|\leq q\leq p,\phantom{X} q\mbox{ even if }p\mbox{ is even},\phantom{X} q\mbox{ odd if }p\mbox{ is odd} \bigg\} 
\end{equation}
and $f_{p;2q,m}\in\Complex$. The functions $\TT{m}{j}{k}$ appearing in
the expansion (\ref{normal expansion}) are the functions of
$SU(2,\Complex)$ onto $\Complex$ given by
\begin{subequations}
\begin{eqnarray}
&& SU(2,\Complex) \ni t\mapsto \TT{m}{j}{k}(t)=\binom{m}{j}^{1/2} \binom{m}{k}^{1/2} t^{(B_1}_{\phantom{(B_1}(A_1}\cdots t^{B_m)_j}_{\phantom{{B_m)_j}}A_m)_k}, \\
&& \TT{0}{0}{0}(t)=1, \quad j,k=0,\ldots,m, \quad m=1,2,3,\ldots,
\end{eqnarray} 
\end{subequations} 
where the string of indices with a lower index $k$, say, means that the
indices are symmetrised and then $k$ of them are set equal to 1, while
the remaining ones are set equal to $0$. The function $f$ will be said to have \emph{axial symmetry} if its expansion (\ref{normal expansion}) is of the form
\begin{equation} 
f=\sum^\infty_{p=|s|} \frac{1}{p!} \sum_{q\in Q_p} f_{p;2q,q} \TT{2q}{q}{q-s} \rho^p.
\end{equation}

The following properties of the functions $\TT{m}{j}{k}$ are crucial:
under complex conjugation they transform as
\begin{equation}
\overline{\TT{m}{j}{k}}=(-1)^{j+k}\TT{m}{m-j}{m-k}.
\end{equation}
Furthermore
\begin{eqnarray*}
&& X_+\TT{m}{k}{j}=\sqrt{j(m-j+1)}\TT{m}{k}{j-1} \quad X_-\TT{m}{k}{j}=-\sqrt{(j+1)(m-j)}\TT{m}{k}{j+1}, \\
&& X\TT{m}{k}{j}=(m-2j)\TT{m}{k}{j}.
\end{eqnarray*}
Also important for our aims is the fact that the set
$\{\sqrt{m+1}\TT{m}{j}{k}\}$ constitutes an orthonormal basis of the
space $L^2(\mu,SU(2,\Complex))$ where $\mu$ is the Haar measure on
$SU(2,\Complex)$.

The functions $\TT{m}{k}{j}$ are closely related to the spherical
harmonics. These are defined as a system of orthogonal functions on
$\mathcal{S}^2$. They can be extended to $\mathcal{S}^3$ as functions
with zero spin weight. From this it follows that they can be expanded
as $Y_{lm}=\sum c_{kj}\TT{2k}{j}{k}$. Indeed, one has
\begin{equation}
{}_sY_{nm}=(-i)^{s+2n-m}\sqrt{\frac{2n+1}{4\pi}}\TT{2n}{n-m}{n-s}.
\end{equation} 

Any analytic function on $SU(2,\Complex)$ can be expanded in terms of
functions $\TT{m}{j}{k}$. In particular, the product
$\TT{2n_1}{k_1}{l_1}\times\TT{2n_2}{k_2}{l_2}$ can be \emph{linearised}
rendering
\begin{eqnarray}
&& \hspace{-1cm}\TT{2n_1}{k_1}{l_1}\times\TT{2n_2}{k_2}{l_2}=\sum_{n=q_0}^{n_1+n_2}(-1)^{n+n_1+n_2} C(n_1,n_1-l_1;n_2,n_2-l_2;n,n_1+n_2-l_1-l_2) \nonumber \\
&&\hspace{-0.5cm} \times C(n_1,n_1-k_1; n_2, n_2-k_2; n, n_1+n_2-k_1-k_2)\times \TT{2n}{n+k_1+k_2-n_1-n_2}{n+l_1+l_2-n_1-n_2}, \label{TtimesT}
\end{eqnarray}
where $q_0=\mbox{max}\{|n_1-n_2|,n_1+n_2-k_1-k_2,n_1+n_2-l_1-l_2\}$ and $C(l_1,m_1;l_2,m_2;l,m)$ denote the Clebsch-Gordan coefficients of $SU(2,\Complex)$ \footnote{Some other notations used in the physics literature are:
\[
C(l_1,m_1;l_2,m_2;l,m)=\langle l_1,l_2; m_1,m_2|l,m\rangle =C(l_1,l_2,l|m_1,m_2,m).
\]}. 

Finally, we note that in occasions we will be in the need of dealing
with fields on $\mathcal{B}_a(i)$ which do not project to analytic fields on
$\mathcal{B}_a(i)$, but still have a well defined spin weight. Let $g$ be one
 of such fields with:
\begin{equation}
g=\sum_{p=0} \frac{1}{p!} g_p\rho^p,
\end{equation}
where
\begin{equation}
g_p=\sum_{q=|s|}^{n(p)} \sum_{m=0}^{2q} g_{p;2q,m}\TT{2q}{m}{q-s},
\end{equation}
then we say that the coefficent $g_p$ has expansion type $n(q)$. 

\bigskip
We finish by mentioning that the components of a given spinor arising
in the decompositions (\ref{rank_2}) and (\ref{rank_4}) in terms of
elementary irreducible spinors have definite spin weights. More
precisely, given the rank 2 spinor $\mu_{AB}$, its components $\mu_y$,
$\mu_x$ and $\mu_z$ have spin weight $-1$, $0$ and $+1$
respectively. While for a rank 4 spinor, $\nu_{ABCD}$, the components
$\nu_2$, $\nu_h$ and $\nu_x$ have spin weight $0$; $\nu_1$, $\nu_z$
have spin weight $+1$; $\nu_3$ and $\nu_y$ have spin weight $-1$;
while $\nu_0$ and $\nu_4$ have spin weight $+2$ and $-2$ respectively.

\section{Expansions of solutions to the momentum constraint}
Using the framework described in the previous section, we proceed now
to discuss the solutions to the conformally flat momentum
constraint. The essential content of this section is not new, but the
presentation is. Other approaches can be found in, for example,
\cite{BeiOMu96,DaiFri01}.

We begin with some fairly standard considerations. We shall refer to
$\Real^3$ endowed in standard (Cartesian) coordinates $\{x^a\}$ with
the metric $-\delta_{ab}$ as to the \emph{Euclidean space}
$\Euclidean^3$. In \cite{DaiFri01} it was shown that any smooth
solution in $\mathcal{B}_a(i)\setminus\{i\}\subset \Euclidean^3$ of
the momentum constraint can be written in the form
\begin{equation}\label{psi_tensor}
\psi^{ab}=\psi^{ab}_P+\psi^{ab}_J+\psi^{ab}_A+\psi^{ab}_Q+\psi^{ab}_\lambda,
\end{equation}
where $\psi^{ab}_P=\O(|x|^{-4})$, $\psi^{ab}_J=\O(|x|^{-3})$,
$\psi^{ab}_A=\O(|x|^{-3})$, and $\psi^{ab}_Q=\O(|x|^{-2})$ correspond,
respectively, to the parts of the second fundamental form associated
with translational, rotational, expansion, and boost conformal Killing
vectors. They all consist of $l=0,1$ spherical harmonics. Let $
n^a=x^a/|x|$, then if $\psi^{ab}_P\neq 0$ the
vector
\begin{equation}
P^a=\frac{1}{8\pi}\lim_{\epsilon\rightarrow 0} \int_{S_\epsilon} |x|^2 \psi_{bc} n^b (2 n^c n^a -\delta^{ca}) dS_\epsilon,
\end{equation}
corresponds to the \emph{ADM linear momentum} of the initial
data. Similarly, if $\psi^{ab}_J\neq 0$ then
\begin{equation}
J^a=\frac{1}{8\pi}\lim_{\epsilon\rightarrow 0}\int_{S_\epsilon}|x| \psi_{bc}n^b \epsilon^{cad}n_d dS_\epsilon,
\end{equation}
is the \emph{ADM angular momentum} of the data. In the last two
expressions $S_\epsilon$ denotes a sphere of radius $\epsilon$ centred
at $i$, $dS_\epsilon$ its volume element, and $n^a$ its unit normal.

The term $\psi^{ab}_\lambda$ in equation (\ref{psi_tensor}) is, on the
other hand, associated with $l\geq 2$ spherical harmonics and, thus,
can be interpreted as providing the higher order momenta content of
the second fundamental form. It consists of derivatives of a certain
spin-weight 2 potential $\eth^2 \lambda$, where $\lambda$ is an
arbitrary complex $C^\infty$ function in $\mathcal{B}_a(i)\setminus\{i\}$, and
$\eth$ denotes the standard NP ``eth'' operator. Important for our
purposes is that if $P^a=0$ and $|x|\lambda\in E^\infty(\mathcal{B}_a(i))$
---in which case $\psi^{ab}_\lambda=\O(|x|^{-2})$--- then
$|x|^8\psi^{ab}\psi_{ab}\in E^\infty(\mathcal{B}_a(i))$ ---see
theorem 15 in reference \cite{DaiFri01}--- so that theorem 1 and 2
apply.

\subsection{Spinorial version of the momentum constraint}
In the space spinor formalism, a symmetric tracefree tensor
$\psi_{\alpha\beta}$ is represented by the totally symmetric spinor
$\psi_{ABCD}=\psi_{(ABCD)}$ living on $\mathcal{C}_a$. The spinorial
and tensorial versions are related in a standard way via
\begin{equation}
\psi_{ABCD}=\sigma^i_{AB}\sigma^j_{CD}\psi_{ij}=\sigma^i_{AB}\sigma^j_{CD}e^\alpha_i e^\beta_j \psi_{\alpha\beta}.
\end{equation}
The symmetries of $\psi_{ABCD}$ together with its tracelessness
---$\psi^{AB}_{\phantom{AB}AB}=0$--- imply a decomposition in terms of
elementary spinors of the form:
\begin{equation}
\psi_{ABCD}=\chi_0\epsilon^0_{ABCD}+\psi_1\epsilon^1_{ABCD}+\psi_2\epsilon^2_{ABCD}+\psi_3\epsilon^3_{ABCD}+\psi_4\epsilon^4_{ABCD}.
\end{equation}  
It can be verified that the components $\psi_0$, $\psi_1$, $\psi_2$,
$\psi_3$ and $\psi_4$ are respectively of spin weight
$2,1,0,-1,-2$. The components of $\psi_{ABCD}$, being those of a spinor
associated with a real spatial tensor, must satisfy in virtue of the
reality condition (\ref{reality condition}), the relations
\begin{equation}
\label{reality condition psi}
\psi_2=\overline{\psi}_2, \quad \psi_1=-\overline{\psi}_3, \quad \psi_0=\overline{\psi}_4.
\end{equation}
The momentum constraint can be rewritten as
\begin{equation}
D^{AB}\psi_{ABCD}=0. \label{spacespinor_mc}
\end{equation}

\subsection{The $l=0,1$ harmonic solutions}
Arguably, the most important solutions to the Euclidean momentum
constraint (\ref{spacespinor_mc}) are those having $l=0,1$
harmonics. The importance of this class lies in the fact that
they are associated to conformal Killing vectors of the Euclidean
space. Because of the spin weights of the components of 
$\psi_{ABCD}$, we can write ---cfr. the expansion (\ref{normal expansion})---
\begin{eqnarray*}
&& \psi_0=0, \\
&& \psi_1= \psi_{1,2,0}T_{2\phantom{0}0}^{\phantom{2}0}+\psi_{1,2,1}T_{2\phantom{1}0}^{\phantom{2}1}+\psi_{1,2,2}T_{2\phantom{2}0}^{\phantom{2}2}, \\
&& \psi_2= \psi_{2,0,0}T_{0\phantom{0}0}^{\phantom{0}0}+\psi_{2,2,0}T_{2\phantom{0}1}^{\phantom{2}0}+\psi_{2,2,1}T_{2\phantom{1}1}^{\phantom{2}1}+\psi_{2,2,2}T_{2\phantom{2}1}^{\phantom{2}2}, \\
&& \psi_3= \psi_{3,2,0}T_{2\phantom{0}2}^{\phantom{2}0}+\psi_{3,2,1}T_{2\phantom{1}2}^{\phantom{2}1}+\psi_{3,2,2}T_{2\phantom{2}2}^{\phantom{2}2},\\
&& \psi_4=0.
\end{eqnarray*}
 The substitution of the above expressions into equation (\ref{spacespinor_mc}) renders a solution which can be written as ---cfr. equation (\ref{psi_tensor})---:
\begin{equation}
\psi_{ABCD}=\psi^A_{ABCD}+\psi^P_{ABCD}+\psi^Q_{ABCD}+\psi^J_{ABCD},
\end{equation}
where the non-vanishing components of $\psi^A_{ABCD}$ are given by
\begin{equation} \label{psi A}
 \psi_2^A=-\frac{A}{\rho^3}T_{0\phantom{0}0}^{\phantom{0}0}.
\end{equation}
Those of $\psi^P_{ABCD}$ are
\begin{eqnarray*}
&& \psi_1^P=\frac{3}{\rho^4}(P_2+iP_1)T_{2\phantom{0}0}^{\phantom{2}0}-\frac{3\sqrt{2}}{\rho^4}P_3T_{2\phantom{1}0}^{\phantom{2}1}-\frac{3}{\rho^4}(P_2-iP_1)T_{2\phantom{2}0}^{\phantom{2}2}, \\
&& \psi_2^P=\frac{9\sqrt{2}}{2\rho^4}(P_2+iP_1)T_{2\phantom{0}1}^{\phantom{2}0}-\frac{9}{\rho^4}P_3T_{2\phantom{1}1}^{\phantom{2}1}-\frac{9\sqrt{2}}{2\rho^4}(P_2-iP_1)T_{2\phantom{2}1}^{\phantom{2}2} \\
&& \psi_3^P=-\frac{3}{\rho^4}(P_2-iP_1)T_{2\phantom{2}2}^{\phantom{2}2}-\frac{3\sqrt{2}}{\rho^4}P_3T_{2\phantom{1}2}^{\phantom{2}1}+\frac{3}{\rho^4}(P_2+iP_1)T_{2\phantom{2}2}^{\phantom{2}0}.
\end{eqnarray*}
While in the case of $\psi_{ABCD}^Q$ one has,
\begin{subequations}
\begin{eqnarray}
&&\psi_1^Q=\frac{3}{\rho^2}(Q_2+iQ_1)T_{2\phantom{0}0}^{\phantom{2}0}-\frac{3\sqrt{2}}{\rho^2}Q_3T_{2\phantom{1}0}^{\phantom{2}1}-\frac{3}{\rho^2}(Q_2-iQ_1)T_{2\phantom{2}0}^{\phantom{2}2}, \label{psi Q1}\\
&&\psi_2^Q=\frac{9\sqrt{2}}{2\rho^2}(Q_2+iQ_1)T_{2\phantom{0}1}^{\phantom{2}0}-\frac{9}{\rho^2}Q_3T_{2\phantom{1}1}^{\phantom{2}1}-\frac{9\sqrt{2}}{2\rho^2}(Q_2-iQ_1)T_{2\phantom{2}1}^{\phantom{2}2} \label{psi Q2}\\
&&\psi_3^Q=-\frac{3}{\rho^2}(Q_2-iQ_1)T_{2\phantom{2}2}^{\phantom{2}2}-\frac{3\sqrt{2}}{\rho^2}Q_3T_{2\phantom{1}2}^{\phantom{2}1}+\frac{3}{\rho^2}(Q_2+iQ_1)T_{2\phantom{2}2}^{\phantom{2}0}. \label{psi Q3}
\end{eqnarray}
\end{subequations}
And finally for $\psi_{ABCD}^J$,
\begin{subequations}
\begin{eqnarray}
&&\psi_1^J=\frac{6}{\rho^3}(-J_1+iJ_2)\TT{2}{0}{0}+\frac{6\sqrt{2}}{\rho^3}iJ_3\TT{2}{1}{0}-\frac{6}{\rho^3}(J_1+iJ_2)\TT{2}{2}{0}, \label{psi J1}\\
&&\psi_2^J=0 \label{psi J2}\\
&&\psi_3^J=\frac{6}{\rho^3}(J_1+iJ_2)\TT{2}{2}{2}-\frac{6\sqrt{2}}{\rho^3}iJ_3\TT{2}{1}{2}-\frac{6}{\rho^3}(-J_1+iJ_2)\TT{2}{0}{2}, \label{psi J3}
\end{eqnarray}
\end{subequations}
In the above expressions $A$, $P_1$, $P_2$, $P_3$, $J_1$, $J_2$, $J_3$, $Q_1$, $Q_2$, $Q_3\in
\Real$. 

\subsection{Solutions with higher harmonics}
The solutions to the Euclidean momentum constraint with harmonics
$l=0,1$ constitute essentially what is known as the Bowen-York Ansatz
\cite{BowYor80}. This Ansatz does not exhaust all the possibilities as
it excludes solutions with higher harmonics. Some initial data sets of
containing these higher harmonics have been recently considered ---see
\cite{DaiLouTak02}. The interest of these higher harmonic terms lies
in the fact that ---as will be discussed in detail in the article
II--- the second fundamental forms of stationary data, and in
particular the Kerr data, generically do contain this kind of terms.
An example of conformally flat initial data with higher order
multipoles can be found in \cite{DaiLouTak02}. In that reference, the
second fundamental form is essentially that of the $t=constant$ slices
of the Kerr spacetime in Boyer-Lindquist coordinates.

Retaking the strategy used for the $l=0,1$ harmonics solutions, and
because of the linearity of the equations we shall consider the
following Ansatz for the solutions of the momentum constraint:
\begin{equation}
\psi_n=\sum_{q\geq 2} \sum_{k=0}^{2q} L_{n;2q,k}(\rho)\TT{2q}{k}{q-2+n},
\end{equation}
where $n=0,\ldots,4$. Substitution of the latter experssion into equation
(\ref{spacespinor_mc}) yields the following system of ordinary
differential equations for the coefficients $L_{1;2q,k}$, $L_{2;2q,k}$ and $L_{3;2q,k}$, $q\geq 2$, $k=0,\ldots, 2q$: 
\begin{subequations}
\begin{eqnarray}
&&\hspace{-5mm} \frac{\d L_{1;2q,k}}{\d\rho} -\frac{1}{3\rho}\sqrt{q(q+1)}L_{2;2q,k}+\frac{3}{\rho}L_{1;2q,k}-\frac{2}{\rho}\sqrt{(q+2)(q-1)}L_{0;2q,k}=0, \label{ode k1} \\
&&\hspace{-5mm} \frac{\d L_{2;2q,k}}{\d\rho} -\frac{3}{4\rho}\sqrt{q(q+1)}L_{3;2q,k}-\frac{3}{4\rho}L_{1;2q,k}\sqrt{q(q+1)} +\frac{3}{\rho}L_{2;2q,k}=0, \label{ode k2}\\
&&\hspace{-5mm} \frac{\d L_{3;2q,k}}{\d\rho} -\frac{1}{3\rho}\sqrt{q(q+1)}L_{2;2q,k}+\frac{3}{\rho}L_{3;2q,k}-\frac{2}{\rho}\sqrt{(q+2)(q-1)}L_{4;2q,k}=0. \label{ode k3}
\end{eqnarray} 
\end{subequations}

Thus, one sees that if one provides the functions
$L_{0;2q,k}=L_{0;2q,k}(\rho)$ and $L_{4;2q,k}=L_{4;2q,k}(\rho)$, then
it is possible to integrate the above equations to obtain
$L_{1;2q,k}$, $L_{2;2q,k}$ and $L_{3;2q,k}$. In order to fulfill the
reality conditions (\ref{reality condition psi}) one has to require
$L_{0;2q,k}=(-1)^{q+k}\overline{L}_{4;2q,k}$ ---then, the reality conditions
of the other components will be automatically satisfied. In
particular, one could take a complex function $\Lambda\in
C^\infty(\mathcal{B}_a\setminus\{i\})$ and require
\begin{equation}
\sum_{q\geq 2} \sum_{k=0}^{2q} L_{0;2q,k}(\rho)=X_+X_+\Lambda, \quad
\sum_{q\geq 2} \sum_{k=0}^{2q} L_{4;2q,k}(\rho)=X_-X_-\Lambda.
\end{equation}
The operator $X_+$  corresponds essentially to the $\eth$ operator of
Newman-Penrose ---see \cite{FriKan00}. The combination $|x|^4\Lambda$
can be identified with the function $\lambda$ in section 4 of
\cite{DaiFri01}. In order to satisfy the hypothesis of theorem 1 and 2 in section 3.1 ---cfr. theorems 15 of \cite{DaiFri01}--- we require that
\begin{equation}
|x|^4\Lambda \in E^\infty(\mathcal{B}_a(i)), \quad \mbox{i.e. } \Lambda=\frac{\mu}{|x|^4} + \frac{\nu}{|x|^3}, \quad \mu,\nu\in C^\infty(\mathcal{B}_a(i)). 
\end{equation}
Note that under the above requirements $X_+X_+\mu=\O(\rho^2)$ and
$X_+X_+\nu=\O(\rho^2)$. Thus, one has
\begin{equation} \label{psi_lambda04}
 \psi_0^\lambda =\frac{1}{\rho^4} \sum_{n=2}^\infty \sum_{q=2}^n \sum_{k=0}^{2q} L_{0,n-4;2q,k} \TT{2q}{k}{q-2} \rho^n, \quad
 \psi_4^\lambda =\frac{1}{\rho^4} \sum_{n=2}^\infty \sum_{q=2}^n \sum_{k=0}^{2q} L_{4,n-4;2q,k} \TT{2q}{k}{q+2} \rho^n.
 \end{equation}
Under these assumptions the solutions to the system (\ref{ode k1})-(\ref{ode k3}) can be readily calculated yielding
\begin{equation} \label{psi_lambda123}
\psi_j^\lambda =\frac{1}{\rho^4} \sum_{n=2}^\infty \sum_{q=2}^n \sum_{k=0}^{2q} L_{j,n-4;2q,k} \TT{2q}{k}{q-2+j} \rho^n, \quad j=1,2,3,
\end{equation}
where
\begin{subequations}
\begin{eqnarray}
&&\hspace{-2cm} L_{1,n-4;2q,k}=\frac{\left( (L_{4,n-4;2q,k}-L_{0,n-4;2q,k})(q+1)q+4L_{0,n-4;2q,k}(n+3)^2 \right)}{(n+3)(2(n+4)(n+2)-(q+2)(q-1))}\sqrt{(q+2)(q-1)}, \label{K1}\\
&&\hspace{-2cm} L_{2,n-4;2q,k}= \frac{3(L_{0,n-4;2q,k}+L_{4,n-4;2q,k})}{2(n+3)^2-q(q+1)}\sqrt{(q+2)(q+1)q(q-1)},\label{K2} \\
&&\hspace{-2cm} L_{3,n-4;2q,k}= \frac{\left( (L_{0,n-4;2q,k}-L_{4,n-4;2q,k})(q+1)q+4L_{4,n-4;2q,k}(n+3)^2 \right)}{(n+3)(2(n+4)(n+2)-(q+2)(q-1))}\sqrt{(q+2)(q-1)}, \label{K3}
\end{eqnarray}
\end{subequations}
and $n=2,3,4,\ldots$. We conclude by summarising,

\begin{assumption} \label{second ff}
The spinor $\psi_{ABCD}$ associated with the second fundamental form
will be assumed to be on $\mathcal{C}_a$ of the form $
\psi_{ABCD}=\psi^A_{ABCD}+\psi^J_{ABCD}+\psi^Q_{ABCD}+\psi_{ABCD}^\lambda$
where $\psi^A_{ABCD}$ is given by equation (\ref{psi A}),
$\psi^Q_{ABCD}$ by equations (\ref{psi Q1})-(\ref{psi Q3}),
$\psi^J_{ABCD}$ by (\ref{psi J1})-(\ref{psi J3}) and
$\psi^\lambda_{ABCD}$ by equations (\ref{psi_lambda04}),
(\ref{psi_lambda123}) and (\ref{K1})-(\ref{K3}).
\end{assumption}

\section{Expansions of solutions to the Hamiltonian constraint}
The Licnerowicz equation (\ref{Licnerowicz}) is a scalar equation
which can be very easily translated into the space spinor language
rendering:
\begin{equation} \label{hamiltonian_spinor}
D^{AB}D_{AB} \vartheta =\frac{1}{8}\psi_{ABCD}\psi^{ABCD}\vartheta^{-7}.
\end{equation}
Throughout this section we assume that $\psi_{ABCD}$ satisfies the assumption \ref{second ff}, and adopt the local parametrisation 
\begin{equation}
\vartheta=\frac{1}{|x|}+W,
\end{equation}
with $W(i)=m/2$, where $m$ is the ADM mass of the initial data set
$(\widetilde{\mathcal{S}},\widetilde{h}_{\alpha\beta},\widetilde{\chi}_{\alpha\beta})$.
The term $1/|x|$ corresponds to the Green function of the flat
Laplacian $\Delta=D^{AB}D_{AB}$. In the case of non-conformally flat
3-geometries the term $1/|x|$ has to be replaced by an expression of
the form $U/|x|$, where the function $U$ can be calculated recursively
by means of the so-called \emph{Hadamard parametrix construction} ---this
again, will be discussed at lenght in part II. Thus, the term $1/|x|$
contains information about the local geometry near infinity. On the
other hand, the function $W$ contains information which is global in
nature ---the mass in particular.

Under the assumptions of theorem \ref{existence Hamiltonian} the
following expansion for the function $W$ follows:
\begin{equation} \label{W expansion}
W=m/2
+\sum_{p=2}^\infty\sum_{q=0}^p\sum_{k=0}^{2q}\frac{1}{p!}w_{p;2q,k}\TT{2q}{k}{q}\rho^p,
\end{equation}
with $w_{p;2q,k}\in\Complex$, and satisfying the conditions
$\overline{w_{p;2q,k}}=(-1)^{q+k}w_{p;2q,2q-k}$ so as to guarantee
that $W$ is real. Substitution of the above into equation
(\ref{hamiltonian_spinor}) allows us to calculate an expansion 
consistent with the Hamiltonian constraint, but which not necessarily
satisfies it. This is already quite different with what happened in
the time symmetric case where any polynomial of degree $N$ of the form
\begin{equation}
W=m/2
+\sum_{p=2}^N\sum_{k=0}^{2q}\frac{1}{p!}w_{p;2p,k}\TT{2p}{k}{p}\rho^p,
\end{equation}
was an actual solution of the (time symmetric, conformally flat)
Hamiltonian constraint
\begin{equation}
D^{AB}D_{AB}\vartheta=0.
\end{equation}
On the other hand, in the case upon consideration in this article, the
coefficients $w_{p;2q,k}$ depend, in general, in a nontrivial way on
the parameters determining the second fundamental form. As an example,
consider a conformally flat slice in the Schwarzschild spacetime which
is not time symmetric. In this case $\psi_{ABCD}=\psi^A_{ABCD}$. It follows then that the Hamiltonian constraint implies
\begin{equation}
\Delta W =\frac{A^2}{48|x|^6(\vartheta_*+W)^7}, \quad \vartheta_*=\frac{1}{|x|}+\frac{m}{2}.
\end{equation}
A solution to the latter equation can formally be written in terms of the Green function of $\Delta$ as:
\begin{equation}
W=\frac{A^2}{192\pi}\int_{\Real^3}\frac{1}{|x'|^6|x-x'|(\vartheta_*+W)^7}d^3x'.
\end{equation}
In this specific case, it is possible to find the explicit form of the expansion ---e.g. by iterations--- up to a given order. It reads
\begin{equation}
W=-\frac{1}{576}A^2\rho^2+\frac{7}{1920}A^2m\rho^4-\frac{7}{1440}A^2m^2\rho^5+\frac{1}{192}A^2m^3\rho^6-\frac{1}{221184}A^2(1080m^4-A^2)\rho^7+\O(\rho^8).
\end{equation}
For more general second fundamental forms, these similar expansions
have been calculated up to order $\O(\rho^8)$ using the scripts
written in the computer algebra system {\tt Maple V}. Because of their
size, these will not be presented here.

\subsection{Some simplifying considerations}

The conformal factor $\vartheta$ and the symmetric tensor
$\psi_{ab}$ obtained by virtue of theorems 1 and 2 contain
some terms which are pure gauge. The freedom remaining in
our setting consists essentially in elements of the conformal
group. In particular we have at our disposition a rotation and a
translation. We shall make duly use of them. Firstly, a rotation can
be used so that the Cartesian coordinates $\{x^a\}$ are such that
the ADM-angular momentum vector $J^a$ is ``alligned with the
positive z-axis'' ---that is $J_1=J_2=0$.

With regard to the translation freedom, this can be used to eliminate
certain dipolar terms from the expansions of the conformal factor
$\vartheta$. Consider the inversion $x^a=y^a/|y|$ mapping
$\mathcal{B}_a(i)$ to the asymptotic end $\Real^3\setminus
\mathcal{B}_{1/a}(0)$, and consider the conformal factor
$\phi=|x|\vartheta$. This last function satisfies, under the
hypothesis of theorems 1 and 2, an equation of the form $\Delta
\phi=\O(|y|^{-5})$. The assumptions of theorems 1 and 2 ---using
standard arguments of potential theory--- lead to
\begin{equation}
\phi=1 + \frac{m}{2|y|} + \frac{d_a y^a}{|y|^3}+ \O\left( \frac{1}{|y|^3}\right), 
\end{equation} 
where $d^a$ is a constant (dipolar) vector\footnote{It is noted that
without the theorems 1 and 2, $\Delta \phi=\O(|y|^{-5})$ would
actually imply that the order of the remainder in the expansion of
$\phi$ is $\O(|x|^{-3}\ln |x|)$ ---see for example lemma A in the
appendix of \cite{SimBei83}.} . Now, consider the translation $z^a=y^a
+ c^a$, where $c^a$ is a constant vector. One has that,
\begin{equation}
\frac{1}{|y|}=\frac{1}{|z|}\left(1-\frac{c_a z^a}{|z|^2}+\O\left(\frac{1}{|z|^2}\right)\right), \quad \frac{1}{|y|^3}=\frac{1}{|z|^3}\left(1+\O\left(\frac{1}{|z|}\right)\right).
\end{equation}
Where it follows that
\begin{equation}
\phi=1 +\frac{m}{2|z|} + \frac{1}{|z|^3}(d_a z^a-m c_a z^a) + \O\left(\frac{1}{|z|^3}\right)
\end{equation}
so that the dipolar term in the expansions can be removed if one sets
$c_a=d_a/m$ ---``centre of mass''. The final step now consists in a
further inversion $z^a=\tilde{x}^a/|x|^2$. Thus, without loss of
generality one can assume that
\begin{equation}
W=m/2+ q_{ab}x^a x^b+O(|x|^3) ,
\end{equation}
where $q_{ab}$ is a ``quadrupolar term''. Summarising,

\begin{assumption} \label{centre of mass}
The Cartesian (normal) coordinates $\{x^a\}$ on $\mathcal{B}_a(i)$ are
chosen so that the only non-vanishing component of the ADM angular
momemtum is $J_3$ ---and we shall write $J=J_3$--- and the expansion
of $\vartheta$ contains no dipolar terms, that is: $W=m/2+\O(|x|^2)$.
\end{assumption}

\section{The conformal propagation equations}
In the sequel, we shall make use of the dynamical formulation of the
Conformal Einstein field equations developed in
\cite{Fri95,Fri98a}. This formulation of the field equations allows
the implementation of a regular \emph{initial value problem near
spacelike infinity}. Besides the Levi-Civita connection
$\widetilde{\nabla}$ of the spacetime
$(\mathcal{M},\widetilde{g}_{\mu\nu})$ arising as the development of
the initial data
$(\widetilde{\mathcal{S}},\widetilde{h}_{\mu\nu},\chi_{\mu\nu})$, we
shall consider two other connections: the Levi-Civita connection
$\nabla$ of a metric $g_{\mu\nu}=\Theta^2\widetilde{g}_{\mu\nu}$ in
the same conformal class of $\widetilde{g}_{\mu\nu}$; and a conformal
connection (Weyl connection)\footnote{ A Weyl connection is a
torsionfree connection for which parallel transport preserves the
causal nature.}  $\widehat{\nabla}$, which is not necessarily
metric. Now, given the connections $\nabla$, $\widetilde{\nabla}$ and
$\widehat{\nabla}$, there are 1-forms $b_\mu$, $f_\mu$ such that
\begin{subequations}
\begin{eqnarray}
&& \widehat{\nabla}-\widetilde{\nabla}=S(b) \quad \widehat{\nabla}-\nabla=S(f), \\&& \nabla-\widetilde{\nabla}=S(b-f)=S(\Theta^{-1}d\Theta),
\end{eqnarray}
\end{subequations}
where for a given 1-form $d_\mu$ the tensor field $S(d)$ is given by
 \begin{equation}
S(d)_{\mu\phantom{\nu}\rho}^{\phantom{\mu}\nu}=\delta^{\nu}_{\mu}d_\rho+\delta^\nu_\rho d_\mu-g_{\mu\rho}g^{\nu\lambda}d_{\lambda}.
\end{equation}
The Conformal
Einstein equations are essentially equations for a frame, the
connection coefficients of $\widehat{\nabla}$, the Ricci tensor of
$\widehat{\nabla}$, and the \emph{rescaled Weyl tensor} of
$g_{\mu\nu}$.

The Conformal Einstein equations written in terms of Weyl connections
allow the implementation of a certain gauge based on the properties
of certain curves called \emph{conformal geodesics} ---conformal Gauss
coordinates. The conformal geodesic equations for a vacuum spacetime
are given by
\begin{subequations}
\begin{eqnarray}
&& \dot{x}^\nu\widetilde{\nabla}_{\nu} \dot{x}^\mu =-2 b_\nu \dot{x}^\nu \dot{x}^\mu + \widetilde{g}_{\nu\lambda}\dot{x}^\nu\dot{x}^\lambda \widetilde{g}^{\mu\sigma}b_{\sigma}, \\
&& \dot{x}^\nu\widetilde{\nabla}_\nu b_\mu = b_\nu\dot{x}^\nu b_\mu -\frac{1}{2} \widetilde{g}^{\nu\lambda} b_\nu b_\lambda \widetilde{g}_{\mu\sigma}\dot{x}^\sigma, \\
&& \dot{x}^\nu\widetilde{\nabla}_\nu e^\mu_j = -b_\nu e^\nu_j \dot{x}^\mu -b_\nu\dot{x}^\nu e^\mu_j + \widetilde{g}_{\nu\lambda}\dot{x}^\nu e_j^\lambda \widetilde{g}^{\mu\sigma}b_\sigma,
\end{eqnarray}
\end{subequations}
where $x^\mu=x^\mu(\tau)$, $b_\mu=b_\mu(\tau)$, and
$\{e^\mu_j=e^\mu_j(\tau)\}_{j=0,\dots,3}$ are, respectively, a curve,
a 1-form and a frame. For $q\in \widetilde{\mathcal{S}}$, these
equations will be supplemented by the initial conditions at $\tau=0$:
\begin{subequations}
\begin{eqnarray}
&& x(0)=q, \quad \dot{x}(0)=e_0 \mbox{ future directed and orthogonal to } \widetilde{\mathcal{S}}, \\
&& \kappa^{-1}\Omega \widetilde{g}(e_j,e_k)=\eta_{jk}, \quad \kappa>0, \\
&& b_\nu \dot{x}^\nu=0, \ b_\mu=\Omega^{-1} \widetilde{\nabla}_\mu \Omega.
\end{eqnarray}
\end{subequations}
The solutions to the conformal geodesic equations with these initial conditions satisfy ---see e.g. \cite{Fri95,Fri03c}---
\begin{equation}
\dot{x}(\tau)=e_0(\tau), \quad \Theta^2\widetilde{g}(e_j,e_k)=\eta_{jk},
\end{equation}
with a conformal factor $\Theta$ given by
\begin{equation} \label{Theta}
\Theta(\tau)=\kappa^{-1}\Omega \left(1-\tau^2\frac{\kappa^2}{\omega^2}\right), \quad \omega=\frac{2\Omega}{\sqrt{|D_\alpha \Omega D^\alpha \Omega|}}. 
\end{equation}
The function $\kappa$ contains the remaining conformal freedom in this
construction. The 1-form
$b_\mu$ is such that
\begin{equation}
\label{d_1form}
d_j(\tau)\equiv \Theta b_\mu e^\mu_j=\left(-2\tau\frac{\kappa\Omega}{\omega^2},\kappa^{-1}e_k(\Omega)\right),
\end{equation}
where in this last equation $j=0,\ldots 3$ and $k=1,2,3$.

\subsection{ The manifold $\mathcal{M}_{a,\kappa}$}
We introduce some useful terminology. Let
$CSL(2,\Complex)=\Real^+\times SL(2,\Complex)$. Further, let
$CSL(\widetilde{\mathcal{M}})$ denote the fibre bundle with fibres
$CSL(2,\Complex)$. Note that $SL(2,\Complex)$ can be regarded in a
natural way as a subgroup of $CSL(2,\Complex)$, thus
$CSL(\widetilde{\mathcal{M}})$ may be identified with the set of spin
frames $\{\kappa \delta_A\}_{A=0,1}$, where $\{\delta_A\}_{A=0,1}\in
SL(\widetilde{\mathcal{M}})$, $\kappa\in\Real^+$. Now,
$\mathcal{C}_a\subset SU(\mathcal{S})\subset SL(\mathcal{S})$, from
where it follows that $\mathcal{C}_a$ is also a submanifold of
$CSL(\widetilde{\mathcal{M}})$. On the bundle
$CSL(\widetilde{\mathcal{M}})$ we shall be considering its solder form
$\sigma^{AA'}$, its $g$-connection form $\omega^A_{\phantom{A}B}$, and
the connection form associated to the $\widetilde{\nabla}$. We write
\begin{equation}
\label{gammas}
\Gamma_{CC'\phantom{A}B}^{\phantom{CC'}A}=\langle \omega^A_{\phantom{A}B},c_{CC'} \rangle, \quad  \widehat{\Gamma}_{CC'\phantom{A}B}^{\phantom{CC'}A}=\langle \omega^A_{\phantom{A}B},c_{CC'} \rangle,
\end{equation}
where the the \emph{connection coefficients}
$\Gamma_{CC'\phantom{A}B}^{\phantom{CC'}A}$ satisfy
\begin{equation}
\Gamma_{CC'AB}=\Gamma_{CC'BA}, \quad \widehat{\Gamma}_{CC'AB}=\Gamma_{CC'AB}+\epsilon_{CA}f_{BC'}.
\end{equation}

Because of the divergent nature of the rescaled Weyl tensor
$d_{ijkl}=\O(|x|^{-3})$, and accordingly of its spinorial counterpart
$\phi_{ABCD}=\O(\rho^{-3})$ ---cfr. equation (\ref{Weyl
divergent})---, we shall be in the need of considering rescalings,
$\delta\mapsto \kappa^{1/2}\delta$, of the spinor dyads over
$\mathcal{B}_a(i)$, in order to obtain quantities which are regular at
$\rho=0$. A choice of the form $\kappa=\rho \kappa^\prime$, where
$\kappa'$ is a smooth function such that $\kappa^\prime(i)=1$, would
do the trick. The restriction of this rescaling defines a
diffeomorphism of $\mathcal{C}_a$ to
$\mathcal{C}_{a,\kappa}=\kappa^{1/2}\mathcal{C}_a$. This
diffeomorphism can be used to carry to $\mathcal{C}_{a,\kappa}$ the
coordinates $\rho, t^{A}_{\phantom{A}B}$ and the vector fields
$\partial_\rho$, $X$, $X_+$ and $X_-$ defined on $\mathcal{C}_a$.

The conformal factor $\Theta$ given by equation (\ref{Theta}) can be
used to define the following sets:
\begin{subequations}
\begin{eqnarray}
&& \mathcal{M}_{a,\kappa}=\left\{ (\tau,q)\phantom{X}|\phantom{X}q\in \mathcal{C}_{a,\kappa},\; -\frac{\omega(q)}{\kappa(q)}\leq \tau \leq \frac{\omega(q)}{\kappa(q)}\right\}, \\
&& \mathcal{I}=\bigg \{(\tau,q)\in \mathcal{M}_a\phantom{X} |\phantom{X} \rho(q)=0,\; |\tau|<1 \bigg\}, \\
&& \mathcal{I}^\pm = \bigg \{ (\tau,q)\in \mathcal{M}_a\phantom{X}|\phantom{X} \rho(q)=0,\; \tau =\pm 1 \bigg \}, \\
&& \mathscr{I}^\pm =\left\{ (\tau,q)\in \mathcal{M}^+_a\phantom{X}|\phantom{X} q\in \mathcal{C}^+_{a,\kappa},\; \tau=\pm\frac{\omega(q)}{\kappa(q)} \right\}.
\end{eqnarray}
\end{subequations}

Let $\sigma^{AA'}$ denote the solder form of $\mathcal{M}^+_{a,\kappa}$. We use it to define the frame fields $c_{AA'}$ via the conditions
\begin{subequations}
\begin{eqnarray}
&&\langle \sigma^{AA'}, c_{BB'} \rangle =\epsilon_{B}^{\phantom{B}A}\epsilon_{B'}^{\phantom{B'}A'}, \\
&& c_{AA'}=c^0_{AA'}\partial_\tau + c^1_{AA'}\partial_\rho + c^+_{AA'}X_+ + c^-_{AA'}X_-.
\end{eqnarray}
\end{subequations}
These fields can be split in the form
\begin{equation}
c_{AA'}=\frac{1}{2}\tau_{AA'}\tau^{CC'}c_{CC'}-\tau^B_{\phantom{B}A'}c_{AB},
\end{equation}
with
\begin{subequations}
\begin{eqnarray}
&& \tau^{AA'}c_{AA'}=\sqrt{2}\partial_\tau, \\
&& c_{AB}=\tau_{(A}^{\phantom{(A}B'}c_{B)B'}=c^0_{AB}\partial_\tau +c^1_{AB}\partial_\rho+c^+_{AB}X_+ +c^-_{AB}X_-,
\end{eqnarray}
\end{subequations}
and $c^0_{AB}=0$ at $\tau=0$.  Again, the functions $\rho$,
$t^A_{\phantom{A}B}$, $\TT{m}{j}{k}$ are extended off
$\mathcal{C}_{a,\kappa}$ by requiring that they remain constant, for
fixed $q$ along the curves $\tau\mapsto (\tau,q)\in
\mathcal{M}^+_{a,\kappa}$.

The spatial part of the 1-form $d_j$ given in equation (\ref{d_1form}) has a spinorial counterpart $d_{AB}=\tau_{(A}^{\phantom{(A}B'}d_{B)B'}$. The latter can be calculated in terms of the functions $U$ and $W$. Namely,
\begin{equation}
\label{d_AB}
d_{AB}=2\rho\left( \frac{Ux_{AB}-\rho D_{AB}U-\rho^2D_{AB}W}{(U+\rho W)^3}\right).\end{equation}

\subsection{The propagation equations}

Using the conformal geodesic gauge and the 2-spinor decomposition, it
can be shown that the extended conformal field equations given in
\cite{Fri95,Fri98a} under the conformal Gauss coordinates described in the
preceeding sections imply propagation equations for:

\begin{itemize}
\item[(i)] The components of the frame $c^\mu_{AB}$, $\mu=0,1,\pm$.

\item[(ii)] The space-spinor connection coefficients of $\nabla$, 
$\Gamma_{ABCD}=\tau_{B}^{\phantom{B}B'}\Gamma_{AB'CD}$. In
\cite{Fri95} it was shown that in the gauge that we are using
\begin{equation}
\tau^{AA'}\widehat{\Gamma}_{AA'BC}=0, \quad \tau^{AA'}f_{AA'}=0.
\end{equation}
The latter imply the decomposition
\begin{equation}
\Gamma_{ABCD}=\frac{1}{\sqrt{2}}\left(\xi_{ABCD}-\chi_{(AB)CD}\right)-\frac{1}{2}\epsilon_{AB}f_{CD},
\end{equation}
where the spinor $\chi_{ABCD}$ agrees on $\mathcal{S}$ with the second
fundamental form of the initial hypersurface.

\item[(iii)] The spinor $\Theta_{AA'BB'}$ associated with the tensor
\begin{equation}
A_{jk}=\frac{1}{2}\widehat{R}_{(jk)}-\frac{1}{12}\eta^{il}\widehat{R}_{il}\eta_{jk}+ \frac{1}{4}\widehat{R}_{[jk]}.
\end{equation}
where $\widehat{R}_{jk}$ is the Ricci tensor of the Weyl
connection $\widehat{\nabla}$.  In the above expression all indices range
$0,\ldots,3$. The spinor $\Theta_{AA'BB'}$ can be decomposed as
\begin{equation}
\Theta_{AA'BB'}=\Phi_{AA'BB'}+\Lambda\epsilon_{AB}\epsilon_{A'B'} +\Phi_{AB}\epsilon_{A'B'}+\overline{\Phi}_{A'B'}\epsilon_{AB}.
\end{equation}
Now, in our gauge
\begin{equation}
\Theta^{BB'}\Theta_{AA'BB'}=0.
\end{equation}
The propagation equations do not contain $\Theta_{AA'BB'}$, but its space spinor counterpart $\Theta_{ACBD}=\tau_{C}^{\phantom{C}A'}\tau_{D}^{\phantom{D}B'}\Theta_{AA'BB'}$. Moreover, $\Theta_{ABC}^{\phantom{ABC}C}=0$, so we can write
\begin{equation}
\Theta_{ABCD}=\Theta_{(AB)CD}-\frac{1}{2}\epsilon_{AB}\Theta_{G\phantom{G}CD}^{\phantom{G}G},
\end{equation}
with $\Theta_{(AB)CD}=\Theta_{(AB)(CD)}$. 

\item[(iv)] The components of the Weyl spinor  $\phi_{ABCD}=\phi_{(ABCD)}$,
where
\begin{equation}
d_{ijkl}=\sigma^{AA'}_{i}\sigma^{BB'}_{j}\sigma^{CC'}_{k}\sigma^{DD'}_{l}\bigg(\phi_{ABCD}\epsilon_{A'B'}\epsilon_{C'D'}+\overline{\phi}_{A'B'C'D'}\epsilon_{AB}\epsilon_{CD}\bigg).
\end{equation}
 
\end{itemize}

The propagation equations group naturally in two sets: the propagation
equations for what will be known as the $v$-quantities,
$v=\left(c^\mu_{AB},\xi_{ABCD},f_{AB},\chi_{(AB)CD},\Theta_{(AB)CD},\Theta_{G\phantom{G}CD}^{\phantom{G}G}
\right)$, $\mu=0,1,\pm$
\begin{subequations}
\begin{eqnarray}
&&\partial_\tau c^0_{AB}=-\chi_{(AB)}^{\phantom{(AB)}EF}c^{0}_{EF}-f_{AB}, \label{p1} \\
&&\partial_\tau c^\alpha_{AB}=-\chi_{(AB)}^{\phantom{(AB)}EF}c^\alpha_{EF}, \label{p2}\\
&&\partial_\tau \xi_{ABCD}=-\chi_{(AB)}^{\phantom{(AB)}EF}\xi_{EFCD}+\frac{1}{\sqrt{2}}(\epsilon_{AC}\chi_{(BD)EF}+\epsilon_{BD}\chi_{(AC)EF})f^{EF} \nonumber\\
&&\hspace{2cm} -\sqrt{2}\chi_{(AB)(C}^{\phantom{(AB)(C}E}f_{D)E}-\frac{1}{2}(\epsilon_{AC}\Theta_{F\phantom{F}BD}^{\phantom{F}F}+\epsilon_{BD}\Theta_{F\phantom{F}AC}^{\phantom{F}F})-i\Theta\mu_{ABCD}, \label{p3} \\
&&\partial_\tau f_{AB}=-\chi_{(AB)}^{\phantom{(AB)}EF}f_{EF}+\frac{1}{\sqrt{2}}\Theta_{F\phantom{F}AB}^{\phantom{F}F}, \label{p4} \\
&&\partial_\tau \chi_{(AB)CD}=-\chi_{(AB)}^{\phantom{(AB)}EF}\chi_{EFCD}-\Theta_{(CD)AB}+\Theta\eta_{ABCD}, \label{p5} \\
&&\partial_\tau\Theta_{(AB)CD}=-\chi_{(AB)}^{\phantom{(AB)}EF}\Theta_{(CD)EF}-\partial_\tau\Theta\eta_{ABCD}+i\sqrt{2}d^E_{\phantom{E}(A}\mu_{B)CDE}, \label{p6} \\
&&\partial_\tau\Theta_{G\phantom{G}AB}^{\phantom{G}G}=-\chi_{(AB)}^{\phantom{(AB)}EF}\Theta_{G\phantom{G}EF}^{\phantom{G}G}+\sqrt{2}d^{EF}\eta_{ABEF}, \label{p7}
\end{eqnarray}
\end{subequations}
where 
\begin{equation}
\eta_{ABCD}=\frac{1}{2}(\phi_{ABCD}+\phi_{ABCD}^+), \quad
\mu_{ABCD}=-\frac{i}{2}(\phi_{ABCD}-\phi_{ABCD}^+),
\end{equation}
denote, respectively, the electric and magnetic parts of
$\phi_{ABCD}$. The quantities $\Theta$, $\partial_\tau\Theta$
correspond to the conformal factor given by equation (\ref{Theta})
awhile $d_{AB}$ is given by equation (\ref{d_AB}). These quantities
are available \emph{a priori} from a knowledge of the solutions to the
constraint equations. Thus, the equations (\ref{p1})-(\ref{p7}) are
essentially ordinary differential equations for the components of the
vector $v$.

The second set of equations is, arguably, the most important part of
the propagation equations and corresponds to the evolution equations
for the spinor $\phi_{ABCD}$ derived from the Bianchi identities
\emph{Bianchi propagation equations}:
\begin{subequations} 
\begin{eqnarray}
&&\hspace{-1.5cm}(\sqrt{2}-2c^0_{01})\partial_\tau\phi_0+2c^0_{00}\partial_\tau\phi_1-2c^\alpha_{01}\partial_{\alpha}\phi_0+2c^\alpha_{00}\partial_\alpha\phi_1 \nonumber \\
&&\hspace{-0.5cm}= (2\Gamma_{0011}-8\Gamma_{1010})\phi_0+(4\Gamma_{0001}+8\Gamma_{1000})\phi_1-6\Gamma_{0000}\phi_2, \label{b0}\\
&&\hspace{-1.5cm}\sqrt{2}\partial_\tau\phi_1-c^0_{11}\partial_\tau\phi_0+c^0_{00}\partial_\tau\phi_2-c^\alpha_{11}\partial_{\alpha}\phi_0+c^\alpha_{00}\partial_{\alpha}\phi_2\nonumber\\
&&\hspace{-0.5cm}=-(4\Gamma_{1110}+f_{11})\phi_0+(2\Gamma_{0011}+4\Gamma_{1100}-2f_{01})\phi_1
+3f_{00}\phi_2-2\Gamma_{0000}\phi_3, \label{b1} \\
&&\hspace{-1.5cm}\sqrt{2}\partial_\tau\phi_2-c^0_{11}\partial_\tau\phi_1+c^0_{00}\partial_\tau\phi_3-c^\alpha_{11}\partial_{\alpha}\phi_1+c^\alpha_{00}\partial_{\alpha}\phi_3\nonumber \\
&&\hspace{-0.5cm}=-\Gamma_{1111}\phi_0-2(\Gamma_{1101}+f_{11})\phi_1+3(\Gamma_{0011}+\Gamma_{1100})\phi_2-2(\Gamma_{0001}-f_{00})\phi_3-\Gamma_{0000}\phi_4, \label{b2}\\
&&\hspace{-1.5cm}\sqrt{2}\partial_\tau\phi_3-c^0_{11}\partial_\tau\phi_2+c^0_{00}\partial_\tau\phi_4-c^\alpha_{11}\partial_{\alpha}\phi_2+c^\alpha_{00}\partial_{\alpha}\phi_4\nonumber \\
&&\hspace{-0.5cm}=-2\Gamma_{1111}\phi_1
-3f_{11}\phi_2+(2\Gamma_{1100}+4\Gamma_{0011}+2f_{01})\phi_3-(4\Gamma_{0001}-f_{00})\phi_4,
\label{b3}\\
&&\hspace{-1.5cm}(\sqrt{2}+2c^0_{01})\partial_\tau\phi_4-2c^0_{11}\partial_\tau\phi_3+2c^\alpha_{01}\partial_\alpha\phi_4-2c^\alpha_{11}\partial_\alpha\phi_3 \nonumber \\
&&\hspace{-0.5cm}=-6\Gamma_{1111}\phi_2+(4\Gamma_{1110}+8\Gamma_{0111})\phi_3
+(2\Gamma_{1100}-8\Gamma_{0101})\phi_4, \label{b4}
\end{eqnarray}
\end{subequations}
where
\begin{equation}
\phi_j=\phi_{(ABCD)_j}, \qquad j=0,\dots,4.
\end{equation}
The subindex $j$ in $(ABCD)_j$ indicates that after
symmetrisation, $j$ indices are to be set equal to $1$. To the
equations (\ref{b0})-(\ref{b4}) we add yet another set of three
equations, also implied by the Bianchi identities which we refer to as 
the \emph{Bianchi constraint equations}
\begin{subequations}  
\begin{eqnarray}
&&\hspace{-1.5cm}c^0_{11}\partial_\tau\phi_0-2c^0_{01}\partial_\tau\phi_1+c^0_{00}\partial_\tau \phi_2 + c^\alpha_{11}\partial_\alpha\phi_0-2c^\alpha_{01}\partial_\alpha\phi_1+c^\alpha_{00}\partial_\alpha\phi_2 \nonumber \\
&&\hspace{-0.7cm}=-(2\Gamma_{(01)11}-4\Gamma_{1110})\phi_0+(2\Gamma_{0011}-4\Gamma_{(01)01}-4\Gamma_{1100})\phi_1
+6\Gamma_{(01)00}\phi_2-2\Gamma_{0000}\phi_3, \label{c1} \\
&&\hspace{-1.5cm}c^0_{11}\partial_\tau\phi_1-2c^0_{01}\partial_\tau\phi_2+c^0_{00}\partial_\tau \phi_3 + c^\alpha_{11}\partial_\alpha\phi_1-2c^\alpha_{01}\partial_\alpha\phi_2+c^\alpha_{00}\partial_\alpha\phi_3 \nonumber \\
&&\hspace{-0.7cm}=\Gamma_{1111}\phi_0-(4\Gamma_{(01)11}-2\Gamma_{1101})\phi_1+3(\Gamma_{0011}-\Gamma_{1100})\phi_2-(2\Gamma_{0001}-4\Gamma_{(01)00})\phi_3\nonumber\\
&&-\Gamma_{0000}\phi_4, \label{c2}\\
&&\hspace{-1.5cm}c^0_{11}\partial_\tau\phi_2-2c^0_{01}\partial_\tau\phi_3+c^0_{00}\partial_\tau \phi_4 + c^\alpha_{11}\partial_\alpha\phi_2-2c^\alpha_{01}\partial_\alpha\phi_3+c^\alpha_{00}\partial_\alpha\phi_4 \nonumber \\
&&\hspace{-0.7cm}=2\Gamma_{1111}\phi_1-6\Gamma_{(01)11}\phi_2+(4\Gamma_{0011}+4\Gamma_{(01)01}-2\Gamma_{1100})\phi_3-(4\Gamma_{0001}-2\Gamma_{(01)00})\phi_4.
\label{c3}
\end{eqnarray}
\end{subequations}

\subsection{The initial data for the conformal propagation equations}
Assume for the moment that one has a solution
$(\Omega,\chi_{\alpha\beta})$ of the constraint equations
(\ref{conformal_Hamiltonian}) and (\ref{conformal_momentum}). How do
we calculate the initial data for the propagation equations
(\ref{p1})-(\ref{p7}) and (\ref{b0})-(\ref{b4})? In order to do this,
one has to make use of the so-called \emph{conformal constraint field
equations} ---see e.g. \cite{Fri95}.

Let $\Theta$ be the conformal factor induced by the F-gauge. We
consider the further conformal rescaling $\Theta\mapsto
\kappa^{-1}\Theta$ where $\kappa$ is a function on $\mathcal{S}$ such that
$\kappa=\kappa' \rho$ with $\kappa'(i)=1$. This conformal rescaling
induces a rescaling in the spinor frame of the form $\delta \mapsto
\kappa^{1/2} \delta$. Using the procedure described in
\cite{Fri95,Fri98a} one arrives to
\begin{subequations}
\begin{eqnarray}
&&\hspace{-1.5cm} \Theta_{ABCD}=-\frac{\kappa^2}{\Omega} D_{(AB}D_{CD)}\Omega + \frac{\kappa^2}{12}\left(r+\chi_{EFGH}\chi^{EFGH}\right) h_{ABCD}+ \frac{\kappa^2}{\sqrt{2}\Omega}\epsilon_{AB}\chi_{CD}^{\phantom{CD}EF}D_{EF}\Omega, \label{initial_1} \\
&&\hspace{-1.5cm} \phi_{ABCD}=\frac{\kappa^3}{\Omega^2}D_{(AB}D_{CD)}\Omega + \frac{\kappa^3}{\Omega}s_{ABCD} + \frac{\kappa^3}{\Omega}\chi_{EF(AB}\chi_{CD)}^{\phantom{CD)}EF}+\sqrt{2}\frac{\kappa^3}{\Omega}D^E_{\phantom{E}(A}\chi_{BCD)E}, \label{phi_initial}\\
&& \hspace{-1.5cm}c^0_{AB}=0, \quad c^1_{AB}=\kappa x_{AB}, \\
&& \hspace{-1.5cm}c^+_{AB}=\kappa\left(\frac{1}{\rho}z_{AB}+\check{c}^+_{AB}\right), \quad c^-_{AB}=\kappa\left(\frac{1}{\rho}y_{AB}+\check{c}^-_{AB}\right), \\
&& \hspace{-1.5cm}\xi_{ABCD}=\sqrt{2}\left( \kappa \gamma_{ABCD} -\frac{1}{2}(\epsilon_{AC}\kappa_{BD}+\epsilon_{BD}\kappa_{AC})\right), \\
&& \hspace{-1.5cm}\chi_{(AB)CD}=\kappa \Omega^2 \psi_{ABCD}, \\
&& \hspace{-1.5cm}f_{AB}=\kappa_{AB}, \label{initial_2}
\end{eqnarray}
\end{subequations}
with
\begin{equation}
\kappa_{AB}=\left( x_{AB}\partial_\rho +\left(\frac{1}{\rho}z_{AB}+\check{c}^+_{AB}\right)X_+ + \left(\frac{1}{\rho}y_{AB}+\check{c}^-_{AB}\right)X_-\right)\kappa.
\end{equation}
In the above expressions it has been assumed that the initial
hypersurface is maximal but not necessarily conformally flat.

\section{Transport equations at $\mathcal{I}$}

The system of equations (\ref{p1})-(\ref{p7}), (\ref{b0})-(\ref{b4}),
and (\ref{c1})-(\ref{c3}) allow us to introduce a special kind of
(asymptotic) expansions in the region of spacetime near null and
spatial infinity. Writing as before 
$v=\left(c^\mu_{AB},\xi_{ABCD},f_{AB},\chi_{(AB)CD},\Theta_{(AB)CD},\Theta_{G\phantom{G}CD}^{\phantom{G}G}\right)$,
$\mu=0,1,\pm$ and
$\phi=\left(\phi_0,\phi_1,\phi_2,\phi_3,\phi_4\right)$, then the
equations (\ref{p1})-(\ref{p7}) can be concisely written as
\begin{equation}
\label{v_equation}
\partial_\tau v=K v+Q(v,v)+L\phi,
\end{equation}
where $K$, $Q$ denote respectively linear and quadratic functions with
constant coefficients, and $L$ denotes a linear function with
coefficients depending on the coordinates through the functions
$\Theta$, $\partial_\tau\Theta$, $d_{AB}$ and such that
$L|_{\mathcal{I}}=0$. For the Bianchi propagation equations one can
write
\begin{equation}
\label{b_equation}
\sqrt{2} E \partial_\tau \phi + A^{AB}c^\mu_{AB}\partial_\mu\phi=B(\Gamma_{ABCD})\phi,
\end{equation}
where $E$ denotes the $(5\times 5)$ unit matrix, $A^{AB}c^\mu_{AB}$
are $(5\times 5)$ matrices depending on the coordinates, and
$B(\Gamma_{ABCD})$ is a linear $(5\times 5)$ matrix valued function  with constant entries of
the connection coefficients $\Gamma_{ABCD}$. On similar lines, the
Bianchi constraint equations (\ref{c1})-(\ref{c3}) can be written as
\begin{equation}
\label{c_equation}
F^{AB}c^\mu_{AB} \partial_\mu\phi = H(\Gamma_{ABCD}),
\end{equation}
where now $F^{AB}c^\mu_{AB}$ denote $(3\times 5)$ matrices, and
$H(\Gamma_{ABCD})$ is a $(3\times5)$ matrix valued function of the
connection with constant entries.

The system (\ref{v_equation})-(\ref{b_equation}) can be formally
evaluated on $\mathcal{I}$ rendering an interior system of equations
which, due to $L|_{\mathcal{I}}=0$ is decoupled, and can be easily
solved. The solutions $(v^{(0)},\phi^{(0)})$ to this system are
universal ---i.e. independent of the initial data--- and can be
regarded as the leading terms in Taylor-like expansions of the form
\begin{equation}
\label{formal_series}
v \sim \sum_{p=0}\frac{1}{p!}v^{(p)}\rho^p, \quad \phi \sim \sum_{p=0}\frac{1}{p!}\phi^{(p)}\rho^p.
\end{equation}
The higher order coefficients in these expansions can be determined
via a recursive procedure. To do so, we differentiate the systems
(\ref{v_equation}), (\ref{b_equation}) and (\ref{c_equation}) $p$
times and then evaluate at $\mathcal{I}$. The resulting equations are
of the form
\begin{eqnarray}
&&\partial_\tau v^{(p)} = Kv^{(p)}
+Q(v^{(0)},v^{(p)})+Q(v^{(p)},v^{(0)}) \nonumber \\
&&\phantom{XXXXXXXX}+\sum_{j=1}^{p-1}\left(
  Q(v^{(j)},v^{(p-j)})+ L^{(j)}\phi^{(p-j)}\right) +
L^{(p)}\phi^{(0)}. \label{v_transport}
\end{eqnarray} 
From the transport Bianchi propagation
equations (\ref{b0})-(\ref{b4}) one gets
\begin{eqnarray}
&&\left( \sqrt{2}E +A^{AB}(c^0_{AB})^{(0)}\right)\partial_\tau\phi^{(p)} +
A^{AB}(c_{AB}^C)^{(0)}\partial_C\phi^{(p)}= B(\Gamma^{(0)}_{ABCD})\phi^{(p)}
\nonumber \\
&&\phantom{XXXXXXXX}+\sum_{j=1}^p
\left(\begin{array}{c} p \\ j
  \end{array}\right)\left(B(\Gamma_{ABCD}^{(j)})\phi^{(p-j)}-A^{AB}(c^\mu_{AB})^{(j)}\partial_\mu
 \phi^{(p-j)}\right), \label{b_transport}
\end{eqnarray}
where $C=\pm$. Similarly, from the Bianchi constraint equations
(\ref{c1})-(\ref{c3}) one obtains
\begin{eqnarray}
&&F^{AB}(c^0_{AB})^{(0)}\partial_\tau\phi^{(p)}+F^{AB}(c^C_{AB})^{(0)}\partial_C\phi^{(p)}=H(\Gamma^{(0)}_{ABCD})\phi^{(p)}
  \nonumber \\
&&\phantom{XXXXXXXX}+\sum_{j=1}^p
\left(\begin{array}{c} p \\ j
  \end{array}\right)
 \left(
   H(\Gamma_{ABCD}^{(j)})\phi^{(p-j)}-F^{AB}(c^\mu_{AB})^{(j)}\partial_\mu\phi^{(p-j)}\right). \label{c_transport}
\end{eqnarray}

We will refer to these equations as to the \emph{transport
equations of order $p$}. An interesting feature of the transport
equations is that their principal part is universal
---i.e. independent of the order $p$. For $p\geq 1$ they are linear
differential equations for the unknowns of order $p$. Furthermore,
note that the subsystem (\ref{v_transport}) consists only of ordinary
differential equations. Now, the expressions
(\ref{initial_1})-(\ref{initial_2}) for the initial data of the
conformal propagation equations can be used to determine the
expansion types of the quantities $v^{(p)}$ and $\phi^{(p)}$ at
$\mathcal{I}^0$. For inital data satisfying assumptions 1-4, it is found
that they have expansion type $p-1$ and $p$ respectively. The
transport equations (\ref{v_transport}) and (\ref{b_transport}) can be
used in turn to show that the expansion type is preserved during the
evolution.

\bigskip
The transport equations (\ref{v_transport}), (\ref{b_transport}) and
(\ref{c_transport}) are decoupled in the following sense: a knowledge
of $v^{(j)}$, $\phi^{(j)}$, $j=0,\ldots,p-1$ together with the initial
data $v^{(p)}|_{\mathcal{I}^0}$ allows us to solve the subsystem
(\ref{v_transport}) to obtain the quantities $v^{(p)}$. With
$v^{(k)}$, $\phi^{(l)}$, $k=0,\ldots,p$ and $l=0,\ldots,p-1$ and the
initial data $\phi^{(p)}$ at hand one could, in principle, solve the
equations (\ref{b_transport}) to get $\phi^{(p)}$. A major
complication in the aforediscussed procedure is that the coefficient
accompaigning the $\tau$-derivative in equation (\ref{b_transport}) is
such that
\begin{equation}
\left( \sqrt{2}E +A^{AB}(c^0_{AB})^{(0)}\right)=\sqrt{2}\mbox{diag}(1-\tau,1,1,1,1+\tau),
\end{equation}
that is, it looses rank at the sets $\mathcal{I}^\pm$ ---in other
words, the system degenerates. Intuitively, one would expect this
degeneracy of the system to leave some sort of imprint on the
behaviour of its solutions on the critical sets,
$\mathcal{I}^\pm$. From a partial differential equations' point of view,
the degeneracy arises from the transversal intersection of a total
characteristic of the Conformal Einstein field equations, the
cylinder $\mathcal{I}$, with standard characteristics ---null
infinity, $\mathscr{I}$.

In order to obtain a detailed analysis of the behaviour of the
solutions to the propagation Bianchi transport equations at $\mathcal{I}^\pm$
one has to make use of the transport equations implied by the Bianchi
constraint equations (\ref{c_transport}). 

The equations (\ref{b_transport})-(\ref{c_transport})
can be, consistently with the spin weight of its diverse terms,
decomposed in terms of the functions $\TT{i}{j}{k}$. In particular,
one has
\begin{equation}
\phi_j^{(p)}=\sum^p_{q=|2-j|}\sum_{k=0}^{2q}a_{j,p;2q,k}\TT{2q}{k}{q-2+j}.
\end{equation}
with $a_{j,p;2q,k}=a_{j,p;2q,k}(\tau)$. The latter strategy
``reduces'' the problem to the analysis of a system of ordinary
differential equations in $\tau$. Note however, that the left hand sides
of the equations (\ref{b_transport}) and (\ref{c_transport}) involve
products of the components of the vectors $v^{(k)}$ and $\phi^{(l)}$
which need to be linearised, that is, one needs to write them as a
linear combination of $\TT{i}{j}{k}$'s according to formula
(\ref{TtimesT}). Now, proceeding sector by sector, one can use the transport
constraint equations to further reduce the problem to the analysis of
a system of two ordinary differential equations for the coefficients
$a_{0,p;2q,k}$ and $a_{4,p;2q,k}$ of $\phi^{(p)}_0$ and
$\phi^{(p)}_4$. The remaining coefficients can be calculated
algebraically from these two. The equations can be written in the form 
\begin{subequations}
\begin{eqnarray}
&& (1-\tau^2)a_0^{\prime\prime}+\big( 4+2(p-1)\tau\big)a^\prime_0+(q+p)(q-p+1)a_0=b_{0,p;q}(\tau), \label{jacobi_0}\\
&& (1-\tau^2)a_4^{\prime\prime}+\big( -4+2(p-1)\tau\big)a^\prime_4+(q+p)(q-p+1)a_4=b_{4,p;q}(\tau), \label{jacobi_4}
\end{eqnarray}
\end{subequations}
where we have written $a_0$ and $a_4$ instead of $a_{0,p;2q,k}$ and
$a_{4,p;2q,k}$, and where the terms $b_{0,p;q}$ and $b_{4,p;q}$ can be
derived from the right hand sides of (\ref{b_transport}) and
(\ref{c_transport}). Their solution can be written as
\begin{equation}
\label{general_solution}
{ a_{0,p;2q,k}(\tau) \choose a_{4,p;2q,k}(\tau)}=X_{p,q}(\tau) \left[ X^{-1}_{p,q}(0){ a_{0,p;2q,k}(0) \choose a_{4,p;2q,k}(0)} + \int_0^\tau X^{-1}_{p,q}(\tau^\prime)B_{p,q}(\tau^\prime) d\tau^\prime \right],
\end{equation}
where $X_{p,q}(\tau)$ denotes the fundamental matrix of the system
(\ref{jacobi_0})- and the vector $B_{p,q}(\tau)$ is to be derived from $b_{0,p;q}$ and $b_{4,p;q}$. The matrix $X_{p,q}(\tau)$ has been given explicitly in terms of Jacobi polynomials. More importantly, 
\begin{equation}
det X_{p,q}(\tau)= f(\tau)(1-\tau^2)^{p-2}
\end{equation}
with $f(\tau)$ a second order polynomial such that $f(\pm1)\neq
0$. This means that the integrand in equation (\ref{general_solution})
will have poles at $\tau=\pm 1$ unless $B_{p,q}(\tau)$ is very
special. The main hurdle to be overcome in this procedure is that the
explicit form of $B_{p,q}(\tau)$ becomes more and more complicated as
the order $p$ of the expansions increases. Thus, in order to gain some
understanding of its structure, one has to make use of computer
algebra methods and calculate explicitly the solutions up to a given
order.

\subsection{A regularity condition}
A detailed analysis of the expansion types of the diverse quantities involved in equations (\ref{b_transport}) and (\ref{c_transport}) reveals that $B_{p,p}(\tau)=0$. Moreover,
\begin{subequations}
\begin{eqnarray}
&&\hspace{-2cm} a_{0,p;2p,k}(\tau)=(1-\tau)^{p+2}(1+\tau)^{p-2}\left( a_{0,p;2p,k}(0)+{\textstyle\frac{(p+1)(p+2)}{4p}}[a_{0,p;2p,k}(0)-a_{4,p;2p,k}(0)]I_+\right), \\
&&\hspace{-2cm} a_{4,p;2p,k}(\tau)=(1+\tau)^{p+2}(1-\tau)^{p-2}\left( a_{4,p;2p,k}(0)+{\textstyle\frac{(p+1)(p+2)}{4p}}[a_{4,p;2p,k}(0)-a_{0,p;2p,k}(0)]I_-\right),
\end{eqnarray}
\end{subequations}
where
\begin{eqnarray}
&&\hspace{-1cm} I_\pm=\int_0^\tau \frac{d\tau'}{(1\pm\tau^\prime)^{p-1}(1\mp\tau^\prime)^{p+3}}=A_*\ln(1-\tau) +\frac{A_{p\pm 2}}{(1-\tau)^{p\pm2}}+\cdots+ \frac{A_1}{1-\tau} \nonumber \\
&& \hspace{5cm}+ B_*\ln(1+\tau)+\frac{B_{p\mp2}}{(1+\tau)^{p\mp2}}+\cdots+\frac{B_1}{1+\tau}+ C,
\end{eqnarray}
where the $A$'s, $B$'s and $C$ are some constants. Hence, the
coefficients $a_{0,p;2p,k}(\tau)$ and $a_{4,p;2p,k}(\tau)$ will not be
smooth at $\tau=\pm1$ unless 
\begin{equation}
\label{regularity_general}
a_{0,p;2p,k}(0)=a_{4,p;2p,k}(0).
\end{equation}
We shall refer to the latter as to a \emph{regularity
condition}. Although the original derivation of this result in
\cite{Fri98a} was given with time symmetric initial data in mind, the
statement holds true even if the data is not time
symmetric\footnote{In \cite{Fri98a} Friedrich, in a \emph{tour de
force}, has shown that the condition
$\phi_{0,p;2p,k}(0)=\phi_{4,p;2p,k}(0)$ for $p=2,3,\ldots$,
$k=0,\ldots,2p$ is equivalent to the vanishing of the Bach tensor and
its symmetrised, tracefree derivatives to all orders at $i$ ---cfr. also with footnote 1 on the present article.}.

Given the aforediscussed regularity condition, the question arises
whether it is possible to rewrite it in terms of the freely
specifiable data available when solving the constraint equations using
the conformal method. This is the content of the following theorem.

\begin{theorem}
For conformally flat data such that
$|x|^8\psi_{\alpha\beta}\psi^{\alpha\beta}\in E^\infty(\mathcal{B}_a(i))$, and
independently of the choice of the function $\kappa$ appearing in the
conformal factor (\ref{Theta}), the following conditions are
equivalent:
\begin{itemize}
\item[(i)] $\phi_{0,p;2p,k}(0)=\phi_{4,p;2p,k}(0)$, $p=2,3,\ldots$, $k=0,\ldots,2p$;
\item[(ii)] $L_{0,p-4;2p,k}+L_{4,p-4;2p,k}=0$, $p=2,3,\ldots$, $k=0,\ldots,2p$. \end{itemize}
\end{theorem}

\medskip
\emph{\bf Proof.} The result follows from looking at formula
(\ref{phi_initial}) and counting arguments regarding the expansion
types of the quantities involved.

\medskip
\emph{\bf Remark.} Note that the regularity condition
(\ref{regularity_general}) is trivially satisfied for time
symmetric, conformally flat data.

The analysis described in the present article will be concerned with
initial data rendering solutions to the transport equations which are
as smooth as possible. Accordingly, we make the further

\begin{assumption}
The second fundamental form satisfies the condition (ii) in theorem 3. 
\end{assumption}

\section{Solutions to the transport equations and obstructions to the smoothness of null infinity}

From the discussion in the previous section it follows that the
origin of the non-smoothness of the solutions of the transport
equations lies in Bianchi subsystem (\ref{b_transport}). In this
section we shall describe in a qualitative fashion the structure of
the solutions to this subsystem for initial data satisfying the
assumptions 1-5. In order to ease our calculations we choose
\begin{equation}
\kappa=\rho,
\end{equation}
any other choice with $\kappa=\rho\kappa'$, where $\kappa'$ is smooth and
such that $\kappa'(i)=1$ would not alter the essence of the results.

\bigskip
 All the results here presented have
been obtained by means of scripts written for the computer algebra
system {\tt Maple V}. In what follows, we shall be systematically using
the decomposition
\begin{equation}
\phi_j^{(p)}=\sum^p_{q=|2-j|}\sum_{k=0}^{2q}a_{j,p;2q,k}\TT{2q}{k}{q-2+j}.
\end{equation}
of the components of the Weyl spinor $\phi_{ABCD}$. We shall use
$\mathcal{Q}(\tau)$ to denote a generic polynomial in $\tau$, while
$\mathcal{P}_k(\tau)$ will denote a generic polynomial of degree $k$
in $\tau$ such that $\mathcal{P}_k(\pm1)\neq0$.  It will be understood
that $\mathcal{Q}(\tau)$ and $\mathcal{P}_k(\tau)$ appearing in 
different equations are in principle unrelated.

\subsection{Lower order solutions}

The first observation that has to be made is that if the assumptions 1-5 are satisfied then the lower order solutions are fully regular. Indeed,

\begin{theorem}
Under assumption 1-5, the solutions of the transport equations for
$p=0,1,2,3,4$ have polynomial dependence in $\tau$. Thus, they extend
smoothly to the sets $\mathcal{I}^\pm$.
\end{theorem}

It should be pointed out that assumption 5 plays a crucial role
here. The original analysis in \cite{Fri98a} suggests that if the regularity condition involving the Cotton-York tensor does not
hold, then logarithmic divergences in the solutions to the transport
equations can appear at order $p=2$.

\subsection{Solutions at order $p=5$}
As it happens in the time symmetric case, the first logarithmic
divergences in the solutions to the transport equations appear at
order $p=5$. More precisely, the solutions to the $p=5$ v-transport
equations are polynomial in $\tau$. On the other hand, the solutions
of the $p=5$ Bianchi transport equations are such that
\begin{equation}
a_{j,5;4,k}(\tau)=\Upsilon_{5;4,k}\bigg( (1-\tau)^{7-j}\mathcal{P}_j(\tau)\ln(1-\tau) +(1+\tau)^{3+j}\mathcal{P}_{4-j}(\tau)\ln(1+\tau) \bigg) +\mathcal{Q}(\tau), \label{order_5}
\end{equation}
with
\begin{eqnarray}
&& \Upsilon_{5;4,0}=18m^2w_{2;4,0}-\frac{3099}{199}\sqrt{6}m^2L_{0,-2,4,0} \nonumber \\
&& \phantom{XXXXX}+\frac{1023}{199}\sqrt{6}m\Big(L_{0,-1;4,0}-L_{4,-1;4,0}\Big)+\frac{224}{199}\sqrt{6}\Big(L_{0,0;4,0}-L_{4,0;4,0}\Big) \label{obs_540}\\
&& \Upsilon_{5;4,1}=18m^2w_{2;4,1}-\frac{3099}{199}\sqrt{6}m^2L_{0,-2,4,1} \nonumber \\
&& \phantom{XXXXX} +\frac{1023}{199}\sqrt{6}m\Big(L_{0,-1;4,1}-L_{4,-1;4,1}\Big)+\frac{224}{199}\sqrt{6}\Big(L_{0,0;4,2}-L_{4,0;4,2}\Big) \\
&& \Upsilon_{5;4,2}=18w_{2;4,2}m^2 +\frac{37602}{199}m J^2-\frac{3099}{199}\sqrt{6}m^2L_{0,-2,4,2}\nonumber \\
&& \phantom{XXXXX}+\frac{1023}{199}\sqrt{6}m\Big(L_{0,-1;4,2}-L_{4,-1;4,2}\Big)+\frac{224}{199}\sqrt{6}\Big(L_{0,0;4,2}-L_{4,0;4,2}\Big) \\
&& \Upsilon_{5;4,3}=18m^2w_{2;4,3}-\frac{3099}{199}\sqrt{6}m^2L_{0,-2,4,3} \nonumber \\
&& \phantom{XXXXX}+\frac{1023}{199}\sqrt{6}m\Big(L_{0,-1;4,3}-L_{4,-1;4,3}\Big)+\frac{224}{199}\sqrt{6}\Big(L_{0,0;4,3}-L_{4,0;4,3}\Big) \\
&& \Upsilon_{5;4,4}=18m^2w_{2;4,4} -\frac{3099}{199}\sqrt{6}m^2L_{0,-2,4,4}\nonumber \\
&& \phantom{XXXXX}+\frac{1023}{199}\sqrt{6}m\Big(L_{0,-1;4,4}-L_{4,-1;4,4}\Big)+\frac{224}{199}\sqrt{6}\Big(L_{0,0;4,4}-L_{4,0;4,4}\Big). \label{obs_544}
\end{eqnarray}
The coefficients for the remaining sectors: $a_{j;5,0,k}(\tau)$, 
$a_{j;5,2,k}(\tau)$, $a_{j;5,6,k}(\tau)$, $a_{j;5,8,k}(\tau)$ and
$a_{j;5,10,k}(\tau)$, $j=0,\ldots,4$, are polynomial in $\tau$. We
refer to the coefficients $\Upsilon_{5;4,k}$ as to the
\emph{quadrupolar obstructions at order $5$}. Note that although the
initial data set is time asymmetric, the obstructions obtained at this
order are time symmetric.

The quadrupolar sector at order $p=5$ is the only one rendering logarithmic divergences. More precisely
\begin{lemma}
\label{regular_p5}
For initial data satisfying the assumptions 1-5, the 
coefficients $a_{j,5;2q,k}$, $j=0,\ldots 4$, $q=0,\ldots,5$, $q\neq 2$,
$k=0,\ldots,2q$ are polynomials in $\tau$.
\end{lemma}

\subsection{Expansions at order $p=6$}
Some experimentation reveals that if one uses the solutions
(\ref{order_5}) to the $p=5$ Bianchi transport equations then
$\ln(1\pm\tau)$ terms will be present in the vector
$v^{(6)}$. Moreover, $\phi^{(6)}$ would also contain, besides the
$\ln(1\pm\tau)$ terms, expressions involving $\ln^2(1+\tau)$,
$\ln(1+\tau)\ln(1-\tau)$ and $\ln^2(1-\tau)$. The discussion of these
non-regular solutions falls beyond the scope of the present
investigation. Hence, in the following calculation we assume that
\begin{equation}
\Upsilon_{5;4,k}=0, \quad k=0,\ldots 4.
\end{equation}
The latter condition could be used to solve for, say, $w_{2
;4,k}$. Now, at first glance it is not fully clear whether it is
possible to have some initial data for which the obstructions
vanish. As already mentioned in section 5, the coefficients
$w_{p;2q,k}$ in the expansions of the function $W$ are of global
nature and may depend in a non-trivial way on the freely specifiable
data. It may be the case that in order to have initial data for which
the obstruction vanishes further restrictions on the free data are
necessary. Indeed, explict calculations render the following
\begin{lemma}
\label{consistency_6}
Consider initial data satisfying the assumptions 1-5. Let
$\Upsilon_{5;4,k}=0$, $k=0,\ldots,4$. Then the $p=6$ Bianchi
constraint transport equations are satisfied if and only if
\begin{equation}
\frac{1023}{199}\sqrt{6}m\Big(L_{0,-1;4,k}-L_{4,-1;4,k}\Big)+\frac{224}{199}\sqrt{6}\Big(L_{0,0;4,k}-L_{4,0;4,k}\Big)-\frac{3099}{199}\sqrt{6}m^2L_{0,-2,4,k}=0,\end{equation}
for $j=0,\ldots,4$. The latter implies that
\begin{equation}
\label{viol_642}
mw_{2;4,2}=-\frac{2089}{199}J^2,
\end{equation}
while the other $w_{2;4,k}$'s vanish.
\end{lemma}

\bigskip
If $\Upsilon_{5;4,k}=0$ and the previous theorem is satisfied, then the solution to the Bianchi transport equations are of the form:
\begin{equation}
\label{order_6}
 a_{j,6;4,k}(\tau)=\Upsilon^+_{6;4,k}(1-\tau)^{8-j}\mathcal{P}_j(\tau)\ln(1-\tau) +\Upsilon^-_{6;4,k}(1+\tau)^{4+j}\mathcal{P}_{4-j}(\tau)\ln(1+\tau) +\mathcal{Q}(\tau),
\end{equation}
with $k=0,\ldots,6$.

These obstructions do not have a counterpart when considering expansions
of time symmetric initial data. We call the coefficients
$\Upsilon^\pm_{6;4,k}$ the \emph{quadrupolar obstructions at order}
$p=6$. Furthermore, the obstructions are asymmetric with regard to
the two disconnected parts of null infinity. That is,
\begin{equation}
\Upsilon_{6;4,k}^+=0 \not\nLeftrightarrow \Upsilon^-_{6;4,k}=0,
\end{equation}
generically.

In order to analyse with more detail the structure of the
obstructions, we shall further assume that the initial data is
\emph{axially symmetric}. In this case, only the obstructions with
$k=2$ are in principle non-vanishing. They are given by
\begin{subequations}
\begin{eqnarray}
&&\Upsilon^+_{6;4,2}=\frac{7722}{7}iJQ_3-\frac{2198208}{6965}J^2-\frac{20817}{14}AJ^2 \nonumber \\
&& \phantom{XXXXX}+\frac{62691}{4816}\sqrt{6}A\bigg( L_{4,-1;4,2}-L_{0,-1;4,2} \bigg) 
-\sqrt{6}\bigg( \frac{7559711}{126420}L_{4,-1;4,2}+\frac{8613019}{126420}L_{0,-1;4,2}\bigg)  \nonumber \\
&& \phantom{XXXXX}+\frac{58}{43}\sqrt{6}A\bigg(L_{4,0;4,2}-L_{0,0;4,2}\bigg)+\sqrt{6}\bigg( \frac{3282401}{27090}L_{4,0;4,2}+\frac{3648769}{27090}L_{0,0;4,2}\bigg) \nonumber \\
&& \phantom{XXXXX}+\frac{144}{13}\sqrt{6}\bigg(L_{4,1;4,2}+L_{0,1;4,2}\bigg),\\
&& \Upsilon^-_{6;4,2}=\frac{7722}{7}iJQ_3+\frac{2198208}{6965}J^2-\frac{20817}{14}AJ^2  \nonumber \\
&& \phantom{XXXXX}+\frac{62691}{4816}\sqrt{6}A\bigg( L_{4,-1;4,2}-L_{0,-1;4,2} \bigg) 
-\sqrt{6}\bigg( \frac{7559711}{126420}L_{0,-1;4,2}+\frac{8613019}{126420}L_{4,-1;4,2}\bigg) \nonumber \\
&& \phantom{XXXXX}+\frac{58}{43}\sqrt{6}A\bigg(L_{4,0;4,2}-L_{0,0;4,2}\bigg)+\sqrt{6}\bigg( \frac{3282401}{27090}L_{0,0;4,2}+\frac{3648769}{27090}L_{4,0;4,2}\bigg) \nonumber\\
&& \phantom{XXXXX}+\frac{144}{13}\sqrt{6}\bigg(L_{4,1;4,2}+L_{0,1;4,2}\bigg).
\end{eqnarray}
\end{subequations}

The asymmetry in the obstructions can be observed in much simpler
situations. Namely, consider initial data sets for which
\begin{equation}
Q_3=L_{0,-1;4,2}=L_{4,-1;4,2}=L_{0,0,4,2}=L_{4,0;4,2}=L_{0,1,4,2}=L_{4,1;4,2}=0.
\end{equation}
In that case the obstructions read
\begin{equation}
\Upsilon^+_{6;4,2}=-\frac{2198208}{6965}J^2-\frac{20817}{14}AJ^2, \quad \Upsilon^-_{6;4,2}=\frac{2198208}{6965}J^2-\frac{20817}{14}AJ^2,
\end{equation}
which clearly cannot be satisfied simultaneously.

\bigskip
Now, assume one has an initial data such that $\Upsilon^+_{6;4,2}=\Upsilon^-_{6;4,2}=0$. One can use these two conditions to solve for $L_{0,-1;4,2}$, $L_{4,-1;4,2}$. The latter implies that
\begin{equation}
L_{0,-2;4,2}=\frac{14}{23}\bigg( L_{0,0;4,2}-L_{4,0;4,2}\bigg)-\sqrt{6}\frac{19096}{4577}J^2.
\end{equation}
The regularity condition (ii) of theorem 3 together with the reality condition (\ref{reality condition psi}) imply that the coefficient $L_{0,-2;4,2}$ is a pure imaginary number. Similarly,  because of the reality conditions, the first term on the right hand side is also pure imaginary, while the second is clearly a real number. This implies that 
\begin{equation}
J=0, \quad w_{2;4,2}=0.
\end{equation}
This result holds true also for initial data without axial symmetry. More precisely, on has the following
\begin{theorem}
Initial data sets satisfying assumptions 1-5, if $\Upsilon_{5;4,k}=0$, $k=0,\ldots 4$ with the consistency condition of lemma \ref{consistency_6} fulfilled, and moreover, with $\Upsilon^\pm_{6;4,k}$ , $k=0,\ldots 4$ are such that
\begin{equation}
J=0, \quad w_{2;4,k}=0, \mbox{ with } k=0,\ldots,4.
\end{equation}
\end{theorem}

\bigskip
Following a pattern already observed in the case of time symmetric
initial data, the sector $a_{j,6,8,k}(\tau)=0$ will also render
obstructions ---which, in this case, are of octupolar nature and time
symmetric. More precisely
\begin{equation}
a_{j,6;6,k}(\tau)=\Upsilon_{6;6,k}\bigg( (1-\tau)^{8-j}\mathcal{P}_{j+1}(\tau)\ln(1-\tau)+(1+\tau)^{4+j}\mathcal{P}_{5-j}(\tau)\ln(1+\tau)\bigg) +\mathcal{Q}(\tau).
\end{equation}
In particular, one has for example that
\begin{eqnarray}
&&\hspace{-15mm}\Upsilon_{6;6,3}=24w_{3,6,3}-\frac{565753248}{82585}iJ^3 \nonumber \\
&&\hspace{-3mm}+\frac{36399}{830}\sqrt{6}iJ\bigg(L_{4,-1,4,2}-L_{0,-1;4,2}\bigg)+\frac{1904}{415}\sqrt{6}iJ\bigg(L_{4,0;4,2}-L_{0,0;4,2}\bigg) \nonumber \\
&&\hspace{-3mm} -\frac{7272}{415}\sqrt{30}L_{0,-1,6,3}+\frac{408}{83}\sqrt{30}\bigg(L_{0,0;6,3}-L_{4,0;6,3}\bigg)+\frac{576}{415}\sqrt{30}\bigg( L_{0,1;6,3}-L_{4,1;6,3}\bigg).
\end{eqnarray}
The other obstructions have a similar structure. These can be used to
solve for the octupolar quantities $w_{3;6,k}$, $k=0,\ldots,6$. In a
similar way as to what happened when setting $\Upsilon_{5;4,k}=0$, it
is to be expected that consistency conditions will arise in the
expansions at order $p=7$. In complete affinity with lemma
\ref{consistency_6} one has the following

\begin{lemma}
For initial data complying with assumptions 1-5 and such that
$\Upsilon_{5;4,k}=0$, $k=0,\ldots,4$, if the consistency conditions of
lemma \ref{consistency_6} are satisfied, then the coefficients
$a_{j,6;2q,k}(\tau)$ with $j=0,\ldots,4$, $q=0,\ldots,6$, $q\neq,2,3$,
$k=0,\ldots, 2q$ are polynomial in $\tau$.
\end{lemma} 

\subsection{Expansions at order $p=7$ and beyond}

The analogue of lemma \ref{consistency_6} for the expansions at order $p=7$ is the following

\begin{lemma}
\label{consistency_7}
Consider initial data conforming with assumptions 1-5. If, moreover,
$\Upsilon_{5;4,k_2}=\Upsilon^\pm_{6;4,k_2}=0$, and
$\Upsilon_{6;6,k_3}=0$ for $k_i=0,\ldots,2i$ and the consistency
condition in lemma \ref{consistency_6} is observed, then the $p=7$
Bianchi constraint equations are satisfied if and only if
\begin{subequations}
\begin{eqnarray}
&& \hspace{-1.5cm}Q_1=Q_2=Q_3=0;\\
&& \hspace{-1.5cm}L_{0,0;4,k}=L_{4,0;4,k}, \mbox{ with } k=0,\ldots,4; \\
&& \hspace{-1.5cm}85\bigg( L_{0,0;6,k}-L_{4,0;6,k}\bigg) +24\bigg(L_{0,1;6,k}-L_{4,1;6,k}\bigg)-303L_{0,-1;6,k}=0, \mbox{ with }k=0,\ldots,6.
\end{eqnarray}
\end{subequations}
\end{lemma}

\textbf{Remark.} The conditions given in the previous lemma imply that 
\begin{equation}
L_{0,-2;4,k_2}=0, \quad w_{3,6,k_3}=0,
\end{equation}
with $k_i=0,\ldots,2i$.

\bigskip
If the consistency requirements given in the previous lemma are
satisfied then it turns out that the sectors $a_{j,7;0,k_0}(\tau)$,
$a_{j,7;2,k_1}(\tau)$, $a_{j,7;10,k_5}(\tau)$, $a_{j,7;12,k_6}(\tau)$ and
$a_{j,7;14,k_7}(\tau)$ have polynomial dependence in $\tau$, while the
sectors $a_{j,7;4,k_2}(\tau)$, $a_{j,7;6,k_3}(\tau)$ and
$a_{j,7;8,k_4}(\tau)$ exhibit a distinctive structure. Namely,
\begin{equation}
a_{j,7;4,k}(\tau)=\Upsilon^+_{7;4,k}(1-\tau)^{9-j}\mathcal{P}_j(\tau)\ln(1-\tau) +\Upsilon^-_{7;4,k}(1+\tau)^{5+j}\mathcal{P}_{4-j}(\tau)\ln(1+\tau) +\mathcal{Q}(\tau),
\end{equation}
with $k=0,\ldots,4$ and the obstructions $\Upsilon^\pm_{7;4,k}$ time
asymmetric in the sense described in the previous subsection. They
depend on $A$, $L_{0,0;4,k}$, $L_{4,0;4,k}$, $L_{0,1;4,k}$,
$L_{4,1;4,k}$, $L_{0,2;4,k}$, $L_{4,2;4,k}$. The conditions
$\Upsilon^+_{7;4,k}=\Upsilon^-_{7;4,k}=0$ can be used to solve for, say,
$L_{0,0;4,k}$, $L_{4,0;4,k}$.

Similarly, one has
\begin{equation}
a_{j,7;6,k}(\tau)=\Upsilon^+_{7;6,k}(1-\tau)^{9-j}\mathcal{P}_{j+1}(\tau)\ln(1-\tau) +\Upsilon^-_{7;6,k}(1+\tau)^{5+j}\mathcal{P}_{5-j}(\tau)\ln(1+\tau) +\mathcal{Q}(\tau),
\end{equation}
with $k=0,\ldots,6$ and $\Upsilon^\pm_{7;6,3}$ also time asymmetric
and depending on $A$, $L_{0,0;6,k}$, $L_{4,0;6,k}$, $L_{0,1;6,k}$,
$L_{4,1;6,k}$, $L_{0,2;6,k}$ and $L_{4,2;6,k}$. 

Finally, whe have
\begin{equation}
a_{j,7;8,k}(\tau)=\Upsilon_{7;8,k}\bigg( (1-\tau)^{9-j}\mathcal{P}_{j+2}(\tau)\ln(1-\tau)+(1+\tau)^{5+j}\mathcal{P}_{6-j}(\tau)\ln(1+\tau)\bigg) +\mathcal{Q}(\tau), 
\end{equation}
for $k=0,\ldots,8$ which renders a time symmetric obstruction
depending on $w_{3,6,k}$, $A$, $L_{0,0;8,k}$, $L_{4,0;8,k}$,
$L_{0,1;8,k}$, $L_{4,1;8,k}$, $L_{0,2;8,k}$ and $L_{4,2;8,k}$. The
condition $\Upsilon_{7;8,k}$ can be used to solve for $w_{3,6,3}$.

As it is to be expected, requiring that the obstructions obtained at
order $p=7$ all vanish gives rise to consistency conditions on the
expansions at order $p=8$. In particular, one has the following

\begin{lemma}
\label{consistency_8}
Assume that the hypothesis and conditions of lemma \ref{consistency_7}
are satisfied. If one has initial data such that
$\Upsilon^\pm_{7,4,k_2}=0$, $\Upsilon^\pm_{7,6,k_3}=0$, and
$\Upsilon_{7,8,k_4}=0$, with $k_i=0,\ldots,2i$ then a necessary
condtion for the existence of solutions to the $p=8$ constraint
equations is that
\begin{equation}
L_{0,-1,6,k}=L_{4,-1,6,k}=0, \mbox{ with }k=0,\ldots,6.
\end{equation} 
\end{lemma}

From the results here presented it is possible to hint a general
pattern in the structure of the solutions to the transport equations
at an arbitrary order. Nevertheless, it must be said that a rigurous
proof of such a pattern, given the intrincacy and length of the
equations, would constitute a remarkable feat which cannot be attained
with the current tools and understanding of the problem.

\subsection{Asymptotic Schwarzschildean data}
The vanishing of the obstructions and the fulfillment of the
associated compatibility conditions greatly restrict the type of
conformally flat initial data under consideration. We summarise the
results of our calculations in the following

\begin{theorem}
\label{summary}
An initial data set $(\mathcal{S},h_{\alpha\beta},\chi_{\alpha\beta})$ for which
\begin{eqnarray*}
&& \Upsilon_{5,4,k_2}=\Upsilon^\pm_{6,4,k_2}=\Upsilon^\pm_{7,4,k_2}=0, \\
&& \Upsilon_{6,6,k_3}=\Upsilon^\pm_{7,6,k_3}=0, \\
&& \Upsilon_{7,8,k_3}=0,
\end{eqnarray*}
with $k_i=0,\ldots,2i$, and for which the consistency conditions in
lemmas \ref{consistency_6}, \ref{consistency_7} and
\ref{consistency_8} are satisfied is such that
\begin{eqnarray*}
&& W=\frac{m}{2}+\O(\rho^4), \\
&& \psi_{ABCD}=-\frac{A}{\rho^3} \epsilon^2_{ABCD}+ \O(1).
\end{eqnarray*}
\end{theorem}

\bigskip
\textbf{Remark.} Initial data sets of the sort discussed in the
previous theorem can rightfully be called Schwarzschildean up to order
$p=3$. Their essential octupolar ($2^3=8$) content vanishes. More
generally, we shall say that an initial data set is Schwarzschildean
to order $p$ near inifinity if
\begin{equation}
W=\frac{m}{2}+\O(\rho^{p+1}), \quad \psi_{ABCD}=-\frac{A}{\rho^3} \epsilon^2_{ABCD}+ \O(\rho^{p-3}).
\end{equation}

\bigskip
That the solutions of the transport equations for the time asymmetric
data discussed in section 5 extend smoothly to the critical sets
$\mathcal{I}^\pm$ can be shown through an induction argument. Because of the
manifest spherical symmetric of the set up, the only non-zero field
quantities are those with spin-weight $0$ ---that is, the only
non-trivial sector is that given by $\TT{0}{0}{0}$. It is not
difficult to show that the fundamental matrix of the system of
v-transport equations (\ref{v_transport}) has entries that are
polynomial in $\tau$ ---this result is also true even without assuming
spherical symmetry. This implies that the solutions of the
$v$-transport equations will be polynomial if the lower order terms
are polynomial. More crucially is that under the current assumptions, the only
non-trivial Bianchi propagation equation is
\begin{equation}
\partial_\tau\phi_2=-\frac{1}{2}\chi_2\phi_2-3\chi_h\phi_2,
\end{equation}
where $\chi_2$ and $\chi_h$ are, respectively, the $\epsilon^2_{ABCD}$
and $h_{ABCD}$ components of $\chi_{ABCD}$ ---see section 3. From here
it also trivially follows that if the lower order terms in, say, the
$p$th-order Bianchi transport equation are polynomial, then also
$\phi^{(p)}_2$ will be polynomial in $\tau$ and, thus, will extend through
the critical sets. This simple argumentation is possible only because
of the spherical symmetry. This is to be contrasted with the case of,
say, the solutions to the transport equations for Kerrian initial
data, for which ---as it will be discussed in paper II--- all the
components of the Weyl spinor $\phi_{ABCD}$ are non-zero. 

To prove that the solutions of the transport equation for the time
asymmetric Schwarzschildean data render series of the form
(\ref{formal_series}) which are solutions to the propagation equations
is a bit more sophisticated. It can be done using techniques similar to
the ones used by Friedrich in section 6 of \cite{Fri98a}. In any case,
th smoothness follows from the analysis of \cite{Fri04}.

\section{Conclusions}
Theorem \ref{summary} and the structure observed in the expansions up
to order $p=7$ readily suggest the following generalisation of the
conjecture given in \cite{Val04a}. Namely,

\begin{conjecture}
The time development of an asymptotically Euclidean, initial data set
which is conformally flat in a neighbourhood $\mathcal{B}_a(i)$ of
infinity and for which $|x|^8\psi_{\alpha\beta}\psi^{\alpha\beta}\in
E^\infty(\mathcal{B}_a(i))$ admits a conformal extension to both
future and past null infinity of class $C^k$, with $k$ a
non-negative integer, if and only if the initial data is
Schwarzschildean to order $p_\bullet$ in $\mathcal{B}_a(i)$, where
$p_\bullet=p_\bullet(k)$ is a non-negative integer. If the development
admits an extension of class $C^\infty$, then the initial data has to
be exactly Schwarzschildean on $\mathcal{B}_a(i)$.
\end{conjecture}

\textbf{Remarks.} Firstly, it is important to notice ---as it follows
from our discussion in section 8--- that the Schwarzschildean data is
not necessarily time symmetric. The condition
$|x|^8\psi_{\alpha\beta}\psi^{\alpha\beta}\in
E^\infty(\mathcal{B}_a(i))$ is used here for convenience. Any
condition which would ensure that the conformal factor arising from
the Licnerowicz equation is of the form
$\vartheta=\hat{\vartheta}/|x|$ with $\vartheta\in E^\infty$ ---a
conformal factor that is not necessarily smooth, but whose expansions
in $\mathcal{B}_a(i)$ do not contain logarithmic terms--- would serve
as well. One could think of generalisations of the conjecture where
one requires that $\hat{\vartheta}\in E^m(\mathcal{B}_a(i))$ for some
given non-negative integer $m$. If the condition is sharp, it is to be
expected that it would generate its own type of obstructions to the
smoothness of null infinity. Finally, we notice that ---as seen from
the results of our calculations--- future and past null infinity could
have different degrees of regularity, say $C^{k_+}$ and
$C^{k_-}$. This extra structure is lost in the conjecture as clearly
$k=\mbox{min}\{k_+,k_-\}$.

\bigskip
A first step toward proving this conjecture would be to show that the
solutions to the $p$th order transport equations extend smoothly to the
critical sets $\mathcal{I}^\pm$ if and only if the initial data is
Schwarzschildean to a certain order $p_\bullet=p_\bullet(p)$. The
calculations shown in this article suggest that at least
$p_\bullet=p-4$. This is to be contrasted with the results in
\cite{Val04a} which suggest that for conformally flat, time symmetric
data the corresponding value ought to be $p_\bullet=p-3$. This result
which should provide sharp conditions could then be used in turn to
prove the necessary existence results required for the conjecture.

\section*{Acknowledgments}
I am very grateful to R. Beig, S. Dain and H. Friedrich for valuable
conversations and encouragement. I am indebted to C. M. Losert for a
careful reading of the manuscript. I wish to thank the Max Planck
Institute f\"ur Gravitationsphysik, Albert Einstein Institute, where
the late stages of this research were carried for its
hospitality. This research is funded by the Austrian FWF (Fonds zur
Forderung der Wissenschaftlichen Forschung) via a Lise Meitner
fellowship (M690-N02 \& M814-N02).


\end{document}